\documentclass[galley1,jgrga]{agu2001}
\usepackage{times}
\usepackage{graphicx}

\def\mbf#1{\mbox{\boldmath ${#1}$}}
\def\Alfven{Alfv\'{e}n~}
\def\Alfvenic{Alfv\'{e}nic~}
\def\lesssim{\; \buildrel < \over \sim \;}
\def\gtrsim{\; \buildrel > \over \sim \;}

\slugcomment{Submitted to JGR}

\titlerunninghead{Solar Winds by Alfv\'{e}n Waves from Photosphere}
\authorrunninghead{T. K. Suzuki \& S. Inutsuka}

\authoraddr{Department of Physics, Kyoto University, Kitashirakawa, 
Kyoto, 606-8502, Japan; stakeru@scphys.kyoto-u.ac.jp}

\begin{document}
\setkeys{Gin}{draft=false}

\title{Solar Winds Driven by Nonlinear Low-Frequency \Alfven Waves from the 
Photosphere : Parametric Study for Fast/Slow Winds and Disappearance of 
Solar Winds}

\author{Takeru K. Suzuki$^1$ \& Shu-ichiro Inutsuka}
\affil{Department of Physics, Kyoto University, Kitashirakawa, 
Kyoto, 606-8502, Japan}
\altaffiltext{1}{JSPS Research Fellow; Present addoress : Graduate School 
of Arts \& Sciences, University of Tokyo, Meguro, Tokyo, 153-8902, Japan}


\begin{abstract}
We investigate how properties of the corona and solar wind in 
open coronal holes depend on properties of magnetic fields and 
their footpoint motions at the surface. 
We perform one-dimensional magnetohydrodynamical (MHD) simulations for 
the heating and the acceleration in coronal holes by low-frequency 
\Alfven waves from the photosphere to 0.3 or 0.1AU.  
We impose low-frequency ($\lesssim 0.05$Hz) transverse fluctuations of the 
field lines at the photosphere with various amplitude, spectrum, and 
polarization in the open flux tubes 
with different photospheric field strength, $B_{r,0}$, and super-radial 
expansion of the cross section, $f_{\rm max}$.   
We find that transonic solar winds are universal consequences. 
The atmosphere is also stably heated up to $\gtrsim 10^6$K 
by the dissipation of the \Alfven waves through compressive-wave generation 
and wave reflection in the cases of the sufficient wave input with 
photospheric amplitude, $\langle dv_{\perp,0} \rangle \gtrsim 0.7$km s$^{-1}$. 
The density, and accordingly the mass flux, of solar winds show a quite 
sensitive dependence on $\langle dv_{\perp,0} \rangle$ because of an 
unstable aspect of the heating by the nonlinear \Alfven waves. 
A case with $\langle dv_{\perp,0} 
\rangle=0.4$km s$^{-1}$ gives $\simeq$50 times smaller mass flux than 
the fiducial case for the fast wind with 
$\langle dv_{\perp,0} \rangle=0.7$km s$^{-1}$;    
solar wind {\it virtually disappears} only if $\langle dv_{\perp,0} \rangle$ 
becomes $\simeq$1/2. 
We also find that the solar wind speed has a positive correlation with 
$B_{r,0}/f_{\rm max}$, which is consistent with recent observations 
by Kojima et al. 
Based on these findings, we show that both fast and slow solar winds 
can be explained by the single process, the dissipation of the low-frequency 
\Alfven waves, with different sets of $\langle dv_{\perp,0} 
\rangle$ and $B_{r,0}/f_{\rm max}$. Our simulations naturally explain the 
observed (i) anti-correlation of the solar wind speed and the coronal 
temperature and (ii) larger amplitude of \Alfvenic fluctuations 
in the fast wind. 
In Appendix, we also explain our implementation of the outgoing boundary 
condition of the MHD waves with some numerical tests.

\end{abstract}

\begin{article}

\section{Introduction}
\label{sec:int}
The corona can be roughly divided into two sectors with respect to 
magnetic field configurations on the solar surface. 
One is closed field regions in which both footpoints of each 
field line are rooted at the photosphere. The other is open field regions 
in which one footpoint is anchored at the photosphere and the other 
is open into the interplanetary space. 
The open field regions roughly coincide with areas called coronal 
holes which are dark in X-rays.  
From the coronal holes, the hot plasma streams out as solar winds.  


The \Alfven wave, generated by the granulations or other surface 
activities, is a promising candidate operating in the heating and 
acceleration of solar winds from coronal holes. 
It can travel a long distance so that the dissipation plays a role 
in the heating of the solar wind plasma as well as the lower coronal plasma, 
in contrast to other processes, such as magnetic reconnection 
events and compressive waves, the heating of which probably concentrates 
at lower altitude. 
High-frequency ($\sim 10^4$Hz) ioncyclotron waves were 
recently suggested for the inferred preferential heating of the perpendicular 
temperature of the heavy ions \citep{am97,kol98}, whereas there is still 
a possibility that the observed temperature anisotropy might be due to a 
line-of-sight effect \citep{rs04}. 
However, the protons which compose the main part of 
the plasma are supposed to be mainly heated by 
low-frequency ($\lesssim 0.1$Hz) components in the magnetohydrodynamical 
(MHD) regime.  This is first because the low-frequency wave is expected 
to have more power, and second because the resonance frequency of 
the proton is higher than those of heavier ions so that the energy of the 
ioncyclotron wave is in advance absorbed by the heavy ions \citep{cra00}. 

The treatment of the wave dissipation is quite complicated even for 
MHD processes. The main reason is that the nonlinearity becomes  
essential because amplitudes
of upgoing waves are inevitably amplified in the stratified atmosphere 
with decreasing density. 
Therefore, numerical simulation is a powerful tool; 
there are several works on the wave propagation and dissipation 
in one-dimensional (1D) \citep{ls96,ks99,ort03} and in 
two-dimensional \citep{od97,od98,ofm04} simulations. 
However, large density contrast amounting more than 15 orders of magnitude 
from the photosphere to the outer heliosphere has prevented one so far 
from carrying out numerical simulation in a broad region from the 
photosphere to the interplanetary space even in 1D modeling. 

In \citet{szi05} (paper I hereafter) we successfully carried out 1D MHD 
simulation with radiative cooling and thermal conduction from the photosphere 
to 0.3 AU. We showed that the coronal heating and the fast solar 
wind acceleration in the coronal hole are natural consequences of 
the transverse footpoint fluctuations of the magnetic field lines. 
Low-frequency ($\lesssim 0.05$Hz) \Alfven waves are generated by the 
photospheric fluctuations \citep{ul96,cb05}. 
The sufficient amount of the energy transmits 
into the corona in spite of attenuation in the chromosphere and the 
transition region (TR), so that
they effectively dissipate to heat and accelerate the coronal plasma.  
Although it is based on the one-fluid MHD approximation, 
it, for the first time, 
self-consistently treats the plasma heating and the propagation of the 
\Alfven waves from the photosphere to the interplanetary space. 
However, we studied only one case for the fast solar wind in paper I.  
It is important to examine how properties 
of the corona and the solar wind are affected by adopting different types of 
the wave injections and/or different magnetic field geometry of the flux 
tubes. 
The purpose of this paper is to study parametrically the heating 
by the low-frequency \Alfven waves. 

In this paper we adopt the one-fluid MHD approximation in 1D flux tubes,  
which is the same as in paper I. 
This assumption is not strictly correct in the solar wind plasma, 
which will be discussed later (\S \ref{sec:lmt}).
However, we think that the MHD approximation is appropriate 
for studies of the average properties of the plasma in the corona and 
the solar wind 
because the plasma particles are sufficiently randomized by the fluctuating 
magnetic fields.

%

\section{Simulation Set-up}
\label{sec:stup}
\begin{figure}
\figurenum{1} 
\begin{center}
\noindent\includegraphics[width=20pc]{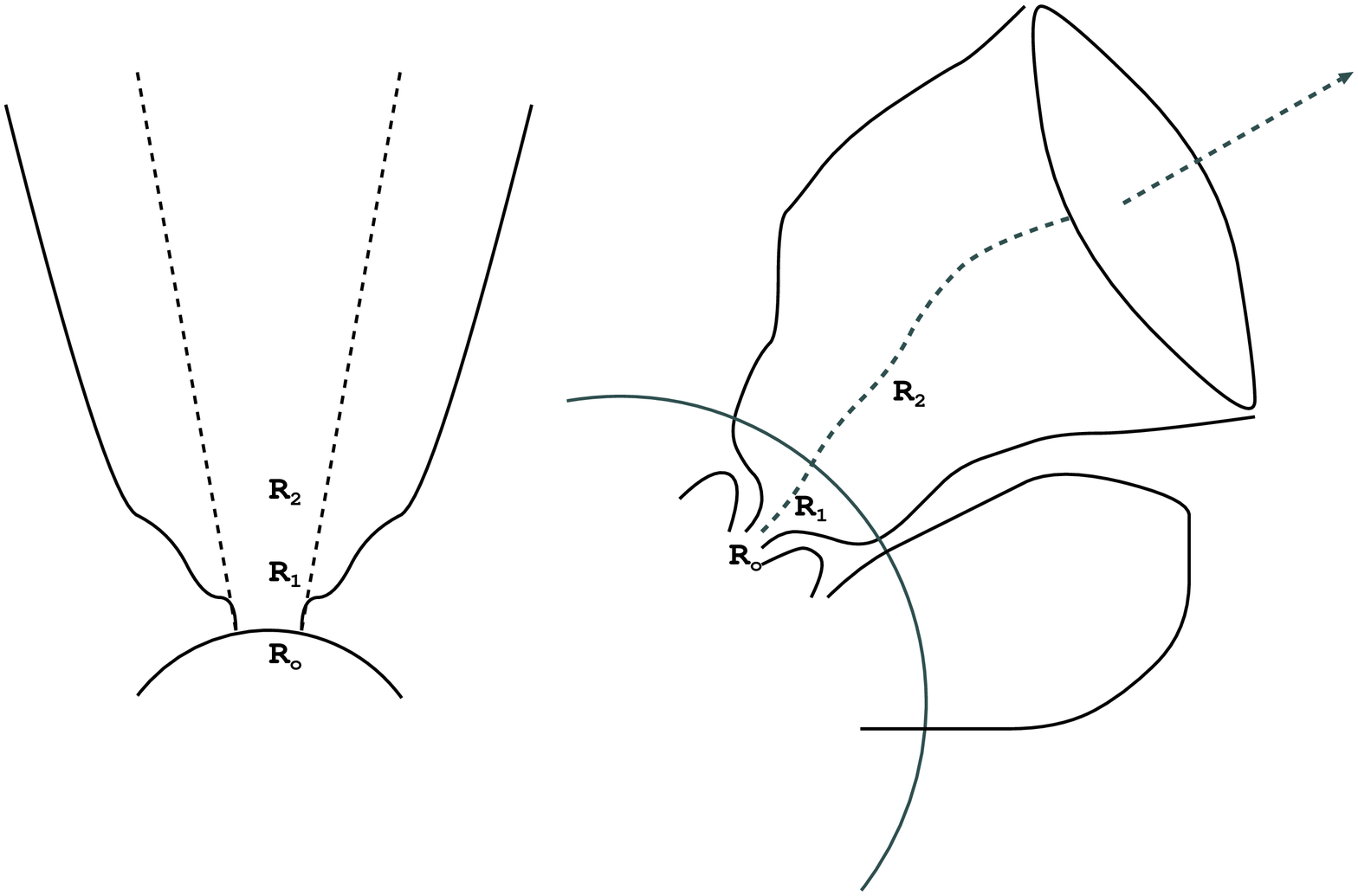}
\end{center}
\caption{
The left figure shows the geometry of the open flux tube in our simulations. 
Arch represents the solar surface and solid lines indicates magnetic field 
lines. Dashed lines correspond to radial expansion. 
The flux tube expands super-radially around $R_1$ and $R_2$. 
This geometry mimics realistic open flux tubes on the sun.  
An example is shown on the right. 
A sizable fraction of the surface is occupied by closed loops with 
a typical length $\sim 10^4$km (=0.014$R_{\odot}$). Above that height 
open structures become more dominant.  Then, 
the open flux tube expands super-radially 
at $\sim R_1$($=1.01R_{\odot}$ in our simulations).  
The tube further expands at $\sim R_2$($=1.2R_{\odot}$ in our simulations) 
due to the large-scale dipole structure.} 
\label{fig:flxdv}
\end{figure}

We consider 1D open flux tubes which are super-radially open, 
measured by heliocentric distance, $r$. 
The simulation regions are from the photosphere ($r=1R_{\odot}$) with  
density, $\rho = 10^{-7}$g cm$^{-3}$, to $65R_{\odot}$ (0.3AU) or $\simeq 
20 R_{\odot}$ ($\simeq$0.1AU), where $R_{\odot}$ is the solar radius.  
Radial field strength, $B_r$, 
is given by conservation of magnetic flux as 
\begin{equation}
B_r r^2 f(r) = {\rm const.} ,
\end{equation}
where $f(r)$ is a super-radial expansion factor. 
We adopt the same function as in \citet{kh76} for $f(r)$ but consider 
two-step expansions: 
$$f(r) = \frac{f_{\rm 1,max}\exp(\frac{r-R_1}{\sigma_1})+f_1}
{\exp(\frac{r-R_1}{\sigma_1})+1} 
\frac{f_{\rm 2,max}\exp(\frac{r-R_2}{\sigma_2})+f_2}
{\exp(\frac{r-R_2}{\sigma_2})+1}$$
where $f_1=1-(f_{\rm 1,max}-1)\exp(\frac{R_{\odot}-R_1}{\sigma_1})$ 
and $f_2=1-(f_{\rm 2,max}-1)\exp(\frac{R_{\odot}-R_2}{\sigma_2})$. 
The flux tube initially expands by a factor of $f_{\rm 1,max}$ around 
$R_1=1.01R_{\odot}$ corresponding to the 'funnel' structure \citep{tu05}, 
and followed by $f_{\rm 2,max}$ times expansion  
around $R_2= 1.2R_{\odot}$ due to the large scale dipole magnetic fields 
(Figure \ref{fig:flxdv}).     
We define the total super-radial expansion as $f_{\rm max} 
=f_{\rm 1,max} f_{\rm 2,max}$.

We prescribe the transverse fluctuations of the field line by the 
granulations at the photosphere, which excite \Alfven waves. 
We study cases of various spectra, polarizations, and root mean squared (rms)
amplitudes, $\langle dv_{\perp,0} \rangle  
(\equiv \sqrt{\langle dv_{\perp,0}^2\rangle}) $(km s$^{-1}$).   
$\langle dv_{\perp,0} \rangle$ 
and power spectrum, $P(\nu)$(erg g$^{-1}$Hz$^{-1}$), have a relation of  
\begin{equation}
\langle dv_{\perp,0}^2 \rangle = 
\int_{\nu_{\rm low\; cut}}^{\nu_{\rm up\; cut}}P(\nu) d\nu . 
\end{equation} 
The parameters adopted in different simulation Runs are tabulated in 
Table 1. 
We consider three types of spectrum : (i) $P(\nu) 
\propto \nu^{-1}$ (ii) $P(\nu) \propto \nu^{0}$ (white noise) 
(iii) $P(\nu) \propto \delta(1/\nu - 180{\rm s})$ (sinusoidal waves with 
a period of 180 seconds) . For the first and second cases  
we set $\nu_{\rm low\; cut} = 6\times 10^{-4}$Hz (a period of 
$\simeq 28$min) and $\nu_{\rm up\; cut}=0.05$Hz (20s) for Run I or 
$\nu_{\rm up\; cut}=0.025$Hz (40s) for the others.  
For the linearly polarized perturbations we only take into account one 
transverse component besides the radial component, while we consider 
the two transverse components for the circularly polarized case. 
Values of $\langle dv_{\perp,0} \rangle$ are chosen to be compatible with  
the observed photospheric velocity amplitude $\sim 1$km s$^{-1}$ 
\citep{hgr78}. 
At the outer boundaries, non-reflecting condition is imposed for all the MHD 
waves (see Appendix), which enables us to carry out 
simulations for a long time until quasi-steady state solutions are obtained  
without unphysical wave reflection.

\begin{table*}[b]
\caption{Model Parameters}
\begin{tabular}{|c||c|c|c|c|c|}
\hline
Run & I & II & III & IV & V  \\
\hline
\hline
Outer Bound.& $65R_{\odot}$ & $20R_{\odot}$   & $20R_{\odot}$  & $20R_{\odot}$   
& $20R_{\odot}$  \\  
Spectrum & $\nu^{-1}$ & $\nu^{-1}$ & 
$\nu^{0}$ & $\delta(\nu^{-1}-180 s)$ & $\delta(\nu^{-1}-180 s)$\\
Polarization & Linear & Linear & Linear & Linear & Circular \\
$\langle dv_{\perp,0} \rangle$(km/s) & 0.7 & 0.7 & 0.7 & 0.7 & 0.7 \\
$B_{r,0}{\rm (G)}$ & $161$ & $161$ & $161$ & $161$ & $161$ \\
$f_{\rm 1,max}$ & 30 & 30 & 30 & 30 & 30 \\
$f_{\rm 2,max}$ & 2.5 & 2.5 & 2.5 & 30 & 2.5 \\
\hline 
Description &Fast Wind&Coarse Ver.& & & \\
 & (paper I) & of Run I & & & \\
\hline
\end{tabular}
\begin{flushright}
\begin{tabular}{|c|c|c|c|c|}
\hline
VI & VII & VIII & IX & X\\
\hline
\hline
$20R_{\odot}$  & $20R_{\odot}$ & $22R_{\odot}$ &
$20R_{\odot}$ & $65R_{\odot}$ \\
$ \nu^{-1}$ & $ \nu^{-1}$& $ \nu^{-1}$ 
& $\nu^{-1}$ & $\nu^{-1}$\\
Linear & Linear & Linear & Linear & Linear\\
1.4 & 0.4 & 0.7 & 0.3 & 1.0\\
$161$ & $161$ & $322$ & $161$ & $322$\\
30 & 30 & 45 & 30 & 45 \\
2.5 & 2.5 & 10 & 2.5 & 10\\
\hline 
 & & &No Steady& Slow Wind\\
 & & & State & \\
\hline
\end{tabular}
\end{flushright}
\end{table*}

We dynamically treat the propagation and dissipation of the waves and the 
heating and acceleration of the plasma by solving 
ideal MHD equations with the relevant physical processes (paper I): 
\begin{equation}
\label{eq:mass}
\frac{d\rho}{dt} + \frac{\rho}{r^2 f}\frac{\partial}{\partial r}
(r^2 f v_r ) = 0 , 
\end{equation}
$$
\rho \frac{d v_r}{dt} = -\frac{\partial p}{\partial r}  
- \frac{1}{8\pi r^2 f}\frac{\partial}{\partial r}  (r^2 f B_{\perp}^2)
$$
\begin{equation}
\label{eq:mom}
+ \frac{\rho v_{\perp}^2}{2r^2 f}\frac{\partial }{\partial r} (r^2 f)
-\rho \frac{G M_{\odot}}{r^2}  , 
\end{equation}
\begin{equation}
\label{eq:moc1}
\rho \frac{d}{dt}(r\sqrt{f} v_{\perp}) = \frac{B_r}{4 \pi} \frac{\partial} 
{\partial r} (r \sqrt{f} B_{\perp}).
\end{equation}
$$
\rho \frac{d}{dt}\left(e + \frac{v^2}{2} + \frac{B^2}{8\pi\rho}
- \frac{G M_{\odot}}{r} \right) 
+ \frac{1}{r^2 f} 
\frac{\partial}{\partial r}\left[r^2 f \left\{ \left(p 
+ \frac{B^2}{8\pi}\right) v_r  \right. \right.
$$
\begin{equation}
\label{eq:eng}
\left. \left.
- \frac{B_r}{4\pi} (\mbf{B \cdot v})\right\}\right]
+  \frac{1}{r^2 f}\frac{\partial}{\partial r}(r^2 f F_{\rm c}) 
+ q_{\rm R} = 0,
\end{equation}
\begin{equation}
\label{eq:ct}
\frac{\partial B_{\perp}}{\partial t} = \frac{1}{r \sqrt{f}}
\frac{\partial}{\partial r} [r \sqrt{f} (v_{\perp} B_r - v_r B_{\perp})], 
\end{equation}
where $\rho$, $\mbf{v}$, $p$, $\mbf{B}$ are density, velocity, pressure, 
and magnetic field strength, respectively, and subscripts 
$r$ and $\perp$ denote radial and tangential components; 
$\frac{d}{dt}$ and $\frac{\partial}{\partial t}$ denote Lagrangian and 
Eulerian derivatives, respectively; 
$e=\frac{1}{\gamma -1}\frac{p}{\rho}$ is specific energy and we assume 
the equation of state for ideal gas with a ratio of specific heat, 
$\gamma=5/3$; 
$G$ and $M_{\odot}$ are the 
gravitational constant and the solar mass; $F_{\rm c}(=\kappa_0 T^{5/2} 
\frac{dT}{dr})$ is thermal conductive flux by Coulomb collisions, where 
$\kappa_0=10^{-6}$ in c.g.s unit \citep{brg65}; 
$q_{\rm R}$ is radiative cooling described below.  
We use optically thin radiative loss\citep{LM90} 
in the corona and upper TR where temperature, 
$T\ge 4\times 10^{4}$K. 
In the chromosphere and low TR, we adopt empirical radiative 
cooling based on the observations \citep{aa89,mor04} 
to take into account the optically thick effect. 
In the low chromosphere, the temperature sometimes drops 
to $\sim 3000$K because we do not consider other heating sources, such as 
sound waves, which must be important there besides \Alfven waves\citep{cs92,
bog03}. 
Such unrealistically low temperature interrupts the propagation of the \Alfven 
waves in the low chromosphere through the change of the density structure. 
To give realistic estimates of the transmission of the \Alfven waves there, 
we switch off the radiative cooling, if the temperature 
becomes $< 5000$K only when $\rho > 10^{-11}$g cm$^{-3}$. 


We adopt the second-order MHD-Godunov-MOCCT scheme (Sano \& Inutsuka 2006) 
to update the physical quantities. 
We solve eqs. (\ref{eq:mass})--(\ref{eq:eng}) on fixed 
Eulerian mesh by remapping the physical variables updated in Lagrangian 
coordinate onto the original grids at each time step.    
The induction equation (\ref{eq:ct}) is solved on the Eulerian 
grids.  
We use implicit time steps for the thermal conduction and radiative 
cooling in energy equation (\ref{eq:eng}) because the Courant conditions 
are severe for these processes, while explicit time steps are adopted 
for the other terms.  
Each cell boundary is treated as discontinuity, and for the time evolution 
we solve nonlinear Riemann shock tube problems with the magnetic pressure term 
by using the Rankin-Hugoniot relations. Therefore, entropy generation, 
namely heating, is automatically calculated from the shock jump condition. 
A great advantage of our code is that no artificial viscosity is required 
even for strong MHD shocks; numerical diffusion is 
suppressed to the minimum level for adopted numerical resolution.  

We initially set static atmosphere with a temperature $T=10^4$K to see 
whether the atmosphere is heated up to coronal temperature and accelerated 
to accomplish the transonic flow. 
At $t=0$ we start the inject of the transverse fluctuations from the 
photosphere and continue the simulations until the quasi-steady states 
are achieved.

\section{Heating and Acceleration by Wave Dissipation}
\label{sec:hawd}
\begin{figure}
\figurenum{2} 
\begin{center}
\noindent\includegraphics[width=20pc]{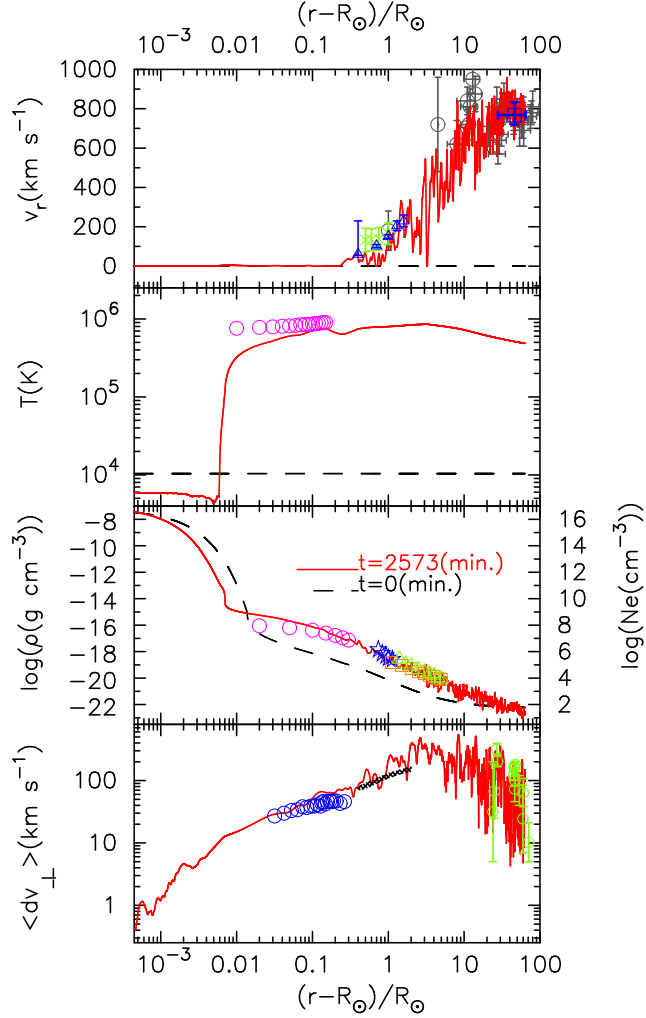}
\end{center}
\caption{Results of Run I with observations of fast solar wind. 
From top to bottom, outflow speed, $v_r$(km s$^{-1}$), temperature, $T$(K), 
density in logarithmic scale, $\log(\rho({\rm g\;cm^{-3}}))$, and rms 
transverse amplitude, $\langle dv_{\perp} \rangle$(km s$^{-1}$) are plotted. 
Observational data in the third panel are electron density, 
$\log(N_e({\rm cm^{-3}}))$ which is to be referred to the right axis. 
Dashed lines indicate the initial conditions and solid lines 
are the results at $t=2573$ minutes. In the bottom panel, the initial 
value ($\langle dv_{\perp} \rangle=0$) dose not appear. 
{\it first}: Green vertical error bars are proton outflow speeds in an
interplume region by UVCS/SoHO \citep{tpr03}. 
Dark blue vertical error bars are proton outflow speeds by the Doppler 
dimming technique using UVCS/SoHO data \citep{zan02}.  
A dark blue open square with errors is velocity by IPS measurements 
averaged in 0.13 - 0.3AU of high-latitude regions \citep{koj04}. 
Light blue data are taken from \citet{gra96}; crossed bars are 
IPS measurements by EISCAT, 
crossed bars with open circles are by VLBA measurements, and 
vertical error bars with open circles are data based on observation 
by SPARTAN 201-01 \citep{hab94}. {\it second}: Pink circles are 
electron temperatures by CDS/SoHO \citep{fdb99}. 
{\it third}: Circles and stars are observations by SUMER/SoHO 
\citep{wil98} and by CDS/SoHO \citep{tpr03}, respectively. 
Triangles \citep{tpr03} and squares \citep{lql97} are observations 
by LASCO/SoHO. 
{\it fourth}: Blue circles are non-thermal broadening inferred from 
SUMER/SoHO measurements \citep{ban98}. Cross hatched region 
is an empirical constraint of non-thermal broadening based on 
UVCS/SoHO observation \citep{ess99}. Green error bars are transverse velocity 
fluctuations derived from IPS measurements by EISCAT\citep{can02}.         
}
\label{fig:obscmp}
\end{figure}

Before showing results of the various coronal holes, 
we explain how the coronal heating and the solar wind acceleration 
were accomplished in the fiducial case (Run I) which was studied
in paper I for the fast solar wind.  
Figure \ref{fig:obscmp} plots the initial condition (dashed lines) and 
the results after the quasi-steady state condition is achieved at $t=2573$ 
minutes (solid lines) in Run I, 
compared with recent observations of fast solar winds. 
From top to bottom, $v_r$(km s$^{-1}$), $T$(K), 
mass density, $\rho$(g cm$^{-3}$), and rms transverse amplitude, 
$\langle dv_{\perp} \rangle$(km s$^{-1}$) are plotted. 
As for the density, we compare our result with observed electron density, 
$N_e$, in the corona. 
When deriving $N_e$ from $\rho$ in the corona, we assume H and He are 
fully ionized, and $N_e({\rm cm^{-3}}) = 6\times 10^{23}\rho$(g cm$^{-3}$). 
These variables are averaged 
for 3 minutes to incorporate observational exposure time. 

Figure \ref{fig:obscmp} shows that the initially cool and static atmosphere 
is effectively heated and accelerated by the dissipation of the \Alfven waves. 
The sharp TR which divides the cool chromosphere with $T\sim 10^4$K and  
the hot corona with $T\sim 10^6$K is formed owing to a thermally unstable 
region around $T\sim 10^5$K in the radiative cooling function \citep{LM90}. 
The hot corona streams out as the transonic solar wind. 
The simulation naturally explains the observed trend quite well. 
(see paper I for more detailed discussions.)

\begin{figure}
\figurenum{3} 
\begin{center}
\noindent\includegraphics[width=20pc]{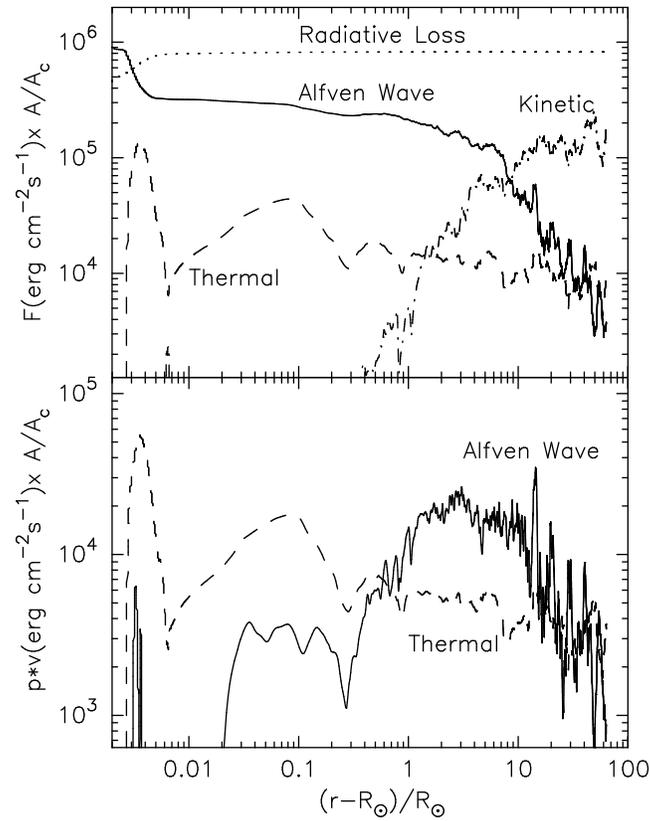}
\end{center}
\caption{Energy flux (top) and momentum flux (bottom) as a function of 
distance from the photosphere. 
Each component is averaged with respect to time during 28min (the longest 
wave period considered). Each value is normalized by cross section, $A_c$, 
of the flow tube at $r=r_c(=1.02R_{\odot})$; 
note that the real flux is larger (smaller) in $r<r_c$ ($r>r_c$).
{\it top}: Solid, dashed, dot-dashed, and dotted lines denote the 
energy flux of the \Alfven wave, the enthalpy flux, the kinetic energy 
flux, and the integrated radiative loss. {\it bottom} Solid and 
dashed lines indicate the advected pressure terms of the \Alfven waves, 
$\langle p_{\rm A} v_r \rangle$ and the thermal pressure, $\langle p v_r 
\rangle$. }
\label{fig:toteng}
\end{figure}

Figure \ref{fig:toteng} shows transfer of the energy (upper) and momentum 
(lower) flux of various components described below.  
The energy equation in an Eulerian form becomes 
$$
\frac{\partial E}{\partial t} + \mbf{\nabla \cdot F} + L_{\rm loss} = 0, 
$$
where $E$ and $\mbf{F}$ are the total energy density and flux. $-L_{\rm loss}$ 
is loss of the energy. 
For the analyses of quasi-steady state behaviors, we can reasonably assume 
that the time-average of the first term equals to 0. 
Then, in our case the explicit form becomes 
$$
\frac{1}{r^2 f}\frac{\partial}{\partial r}\left[r^2 f\left\langle \rho v_r 
\left(\frac{v_r^2 + v_{\perp}^2}{2} + e + \frac{p}{\rho} - 
\frac{G M_{\odot}}{r}\right) \right. \right.
$$
\begin{equation}
\left. \left. 
+ \frac{B_{\perp}^2}{4\pi}v_r -\frac{B_r B_{\perp} v_{\perp}}{4\pi} + 
F_{\rm c} \right\rangle\right] + \langle q_{\rm R}\rangle = 0 ,
\label{eq:engflx}
\end{equation}
where $\langle \cdot \cdot \cdot \rangle$ denotes the time-average.
Using the relation, $v_{\perp} =- B_{\perp}/\sqrt{4\pi\rho}$, for the 
outgoing \Alfven wave, we can extract the energy flux of the \Alfven wave 
from eq.(\ref{eq:engflx}) as follows \citep{jaq77}:
$$
\langle F_{\rm A}\rangle = \langle \rho v_{\perp}^2 (v_{\rm A} + v_r) 
+ p_{\rm A} v_r\rangle = \left\langle -\frac{B_r B_{\perp} v_{\perp}}{4\pi}
\right\rangle 
$$
\begin{equation}
\label{eq:alflx}
+ \left\langle \left(\frac{\rho v_{\perp}^2}{2} 
+ \frac{B_{\perp}^2}{8\pi}\right)v_r \right\rangle 
+ \left\langle  \frac{B_{\perp}^2}{8\pi}v_r \right\rangle , 
\end{equation}
where we 
define \Alfven wave pressure as $p_{\rm A} = B_{\perp}^2/8\pi$. 
In the upper panel of Figure \ref{fig:toteng} we plot the wave energy flux,  
$\langle F_{\rm A} \rangle \frac{r^2 f(r)}{r_c^2 f(r_c)}$,  normalized by 
cross section of the flux tube at $r_c = 1.02 R_{\odot}$ 
with kinetic energy flux, $\rho v_r \frac{v_r^2}{2} \frac{r^2 f(r)}
{r_c^2 f(r_c)}$, enthalpy flux, $\rho v_r (e + \frac{p}{\rho})
\frac{r^2 f(r)}{r_c^2 f(r_c)}$, and integrated radiative loss, 
$\int_{R_{\odot}}^{r} q_{\rm R} \frac{r^2 f(r)}{r_c^2 f(r_c)} dr$. 
Note that the real energy flux is larger (smaller) in $r < r_c$ ($r > r_c$).
While the most of the initial wave energy escapes by radiation 
loss in the chromosphere and the low corona, the remained energy, 
which is $\sim 10\%$ of the input, is transfered to the kinetic energy of the 
solar wind in the outer region.  The thermal energy is almost constant 
in the corona, which indicates that the $\sim$1MK plasma is maintained 
by the energy balance among wave heating, radiative cooling,  
adiabatic cooling (solar wind acceleration), and redistribution by  
thermal conduction. 
 
The lower panel of Figure \ref{fig:toteng} plots measure of pressure 
of \Alfven waves, $\langle p_{\rm A} v_r\rangle $, and thermal gas, 
$\langle p v_r\rangle$. 
This shows that the \Alfven wave pressure dominates the gas pressure 
in the solar wind acceleration region ($1.5R_{\odot} \lesssim r 
\lesssim 10 R_{\odot}$); the fast solar wind is driven by the wave pressure 
rather than by the thermal pressure.  

\begin{figure}
\figurenum{4} 
\begin{center}
\noindent\includegraphics[width=20pc]{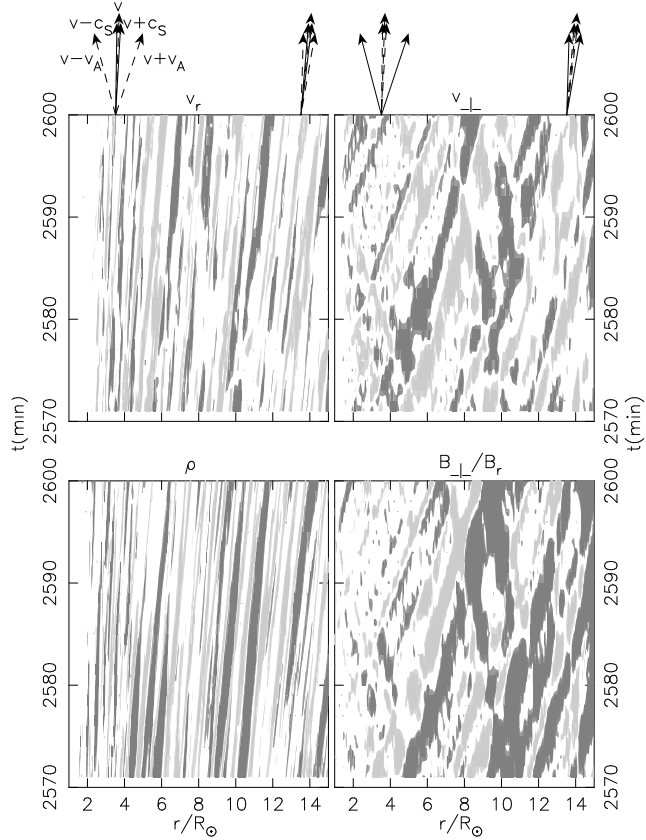}
\end{center}
\caption{$r-t$ diagrams for $v_r$ (upper-left), $\rho$ (lower-left), 
$v_{\perp}$ (upper-right), and $B_{\perp}/B_r$ (lower-right.) 
The horizontal axises cover from $R_{\odot}$ to $15R_{\odot}$, and 
the vertical axises cover from $t=2570$ minutes to $2600$ minutes.  
Dark and light shaded regions indicate positive and negative amplitudes 
which exceed certain thresholds. The thresholds are $d v_r 
=\pm 96$km/s for $v_r$, $d\rho /\rho=\pm0.25$ for $\rho$, 
$v_{\perp}=\pm 180$km/s for $v_{\perp}$, and $B_{\perp}/B_r=\pm 
0.16$ for $B_{\perp}/B_r$, where $d \rho$ and $d v_r$ are 
differences from the averaged $\rho$ and $v_r$. 
Arrows on the top panels indicate characteristics of \Alfven, slow MHD 
and entropy waves at the respective locations (see text). }
\label{fig:rtdgr}
\end{figure}

In paper I, we claimed that the dissipation of the low-frequency \Alfven 
waves in the corona is mainly by the generation of slow waves and 
reflection of the \Alfven waves due to the density fluctuation of the slow 
modes. This can be seen in $r-t$ diagrams. Figure \ref{fig:rtdgr} 
presents contours of amplitude of $v_r$, $\rho$, $v_{\perp}$, and 
$B_{\perp}/B_r$ in $R_{\odot} \le r \le 15 R_{\odot}$ 
from $t=2570$ min. to $2600$ min. 
Dark (light) shaded regions denote positive (negative) amplitude. Above 
the panels, we indicate the directions of the local 5 characteristics, two 
\Alfven, two slow, and one entropy waves at the respective positions. 
Note that the fast MHD and \Alfven 
modes degenerate in our case (wave vector and underlying magnetic field are 
in the same direction), so we simply call the innermost and outermost waves 
\Alfven modes.  
In our simple 1D geometry, $v_r$ and $\rho$ trace the slow modes 
which have longitudinal wave components, while $v_{\perp}$ and $B_{\perp}$ 
trace the \Alfven modes which are transverse (see Cho \& Lazarian [2002;2003] 
for more general cases.). 

One can clearly see the \Alfven waves in $v_{\perp}$ and $B_{\perp}/B_r$ 
diagrams, which have the same slopes with the \Alfven characteristics shown 
above. 
One can also find the incoming modes propagating from lower-right to 
upper-left as well as the outgoing modes generated from the surface\footnote{ 
It is instructive to note that the incoming \Alfven waves have the positive 
correlation between $v_{\perp}$ and $B_{\perp}$ (dark-dark or light-light 
in the figures), while the outgoing modes have the negative correlation 
(dark-light or light-dark).}.  
These incoming waves are generated by the reflection at the `density mirrors'  
of the slow modes (see paper I).
At intersection points of the outgoing and incoming characteristics 
the non-linear wave-wave interactions take place, which play a role 
in the wave dissipation. 

The slow modes are seen in $v_r$ and $\rho$ diagrams. Although it might 
be difficult to distinguish, the most of the patterns are due to the outgoing 
slow modes\footnote{The phase correlation of the longitudinal slow 
waves is opposite to that of the transverse \Alfven waves. 
The outgoing slow modes 
have the positive correlation between amplitudes of $v_r$ and $\rho$, 
($\delta v_r \delta \rho > 0$), while the incoming modes have the negative 
correlation ($\delta v_r \delta \rho < 0$).}
which are generated from the perturbations of the \Alfven wave 
pressure, $B_{\perp}^2/8\pi$ \citep{ks99,tsu02}. 
These slow waves steepen eventually and lead to the shock dissipation. 

The processes discussed here are the combination of the direct mode conversion 
to the compressive waves and the parametric decay instability due to 
three-wave (outgoing \Alfven, incoming \Alfven, and outgoing slow waves) 
interactions (Goldstein 1978; Terasawa et al. 1986; see also \S\ref{sec:wpl}) 
of the \Alfven waves. 
Although they are not generally efficient in the homogeneous background 
since they are the nonlinear mechanisms, the density gradient of the 
background plasma totally changes the situation. 
Owing to the gravity, the density rapidly decreases in the corona as $r$ 
increases, which results in the amplification of the wave amplitude so that 
the waves easily become nonlinear. 
Furthermore, the \Alfven speed varies a lot due to the change of the density 
even within one wavelength of \Alfven waves with periods of minutes or longer. 
This leads to both variation of the wave pressure in one wavelength  
and partial reflection through the deformation of the wave shape 
\citep{moo91}.
The dissipation is greatly enhanced by the density stratification, 
in comparison with the case of the homogeneous background. 
Thus, the low-frequency \Alfven waves are effectively dissipated,  
which results in the heating and acceleration of the coronal plasma. 
 

In summary, we have shown that the low-frequency \Alfven waves are 
dissipated via the combination of the direct mode conversion and the decay 
instability which are considerably enhanced by the long wavelength 
(so called ``non-WKB'') effect in the stratified atmosphere. 
Although each process is not individually effective enough, 
the significant dissipation rate is realized by the combination.   

\section{Universality and Diversity}
We investigate how the properties of the corona and solar wind depend 
on the parameters of the input photospheric fluctuations 
and the flux tube geometry.   
We use smaller simulation boxes in Run II-IX 
than in Run I (Table 1) 
to save the computational time for the parameter studies.  
In most cases quasi-steady state conditions are achieved after $t\gtrsim 
30$hr. In this section we show results at $t=48.3$hr unless 
explicitly stated. 

\subsection{Various Input Waves}
\label{sec:viw}

\subsubsection{Wave Spectrum}
\label{sec:wsp}
\begin{figure}
\figurenum{5} 
\begin{center}
\noindent\includegraphics[width=20pc]{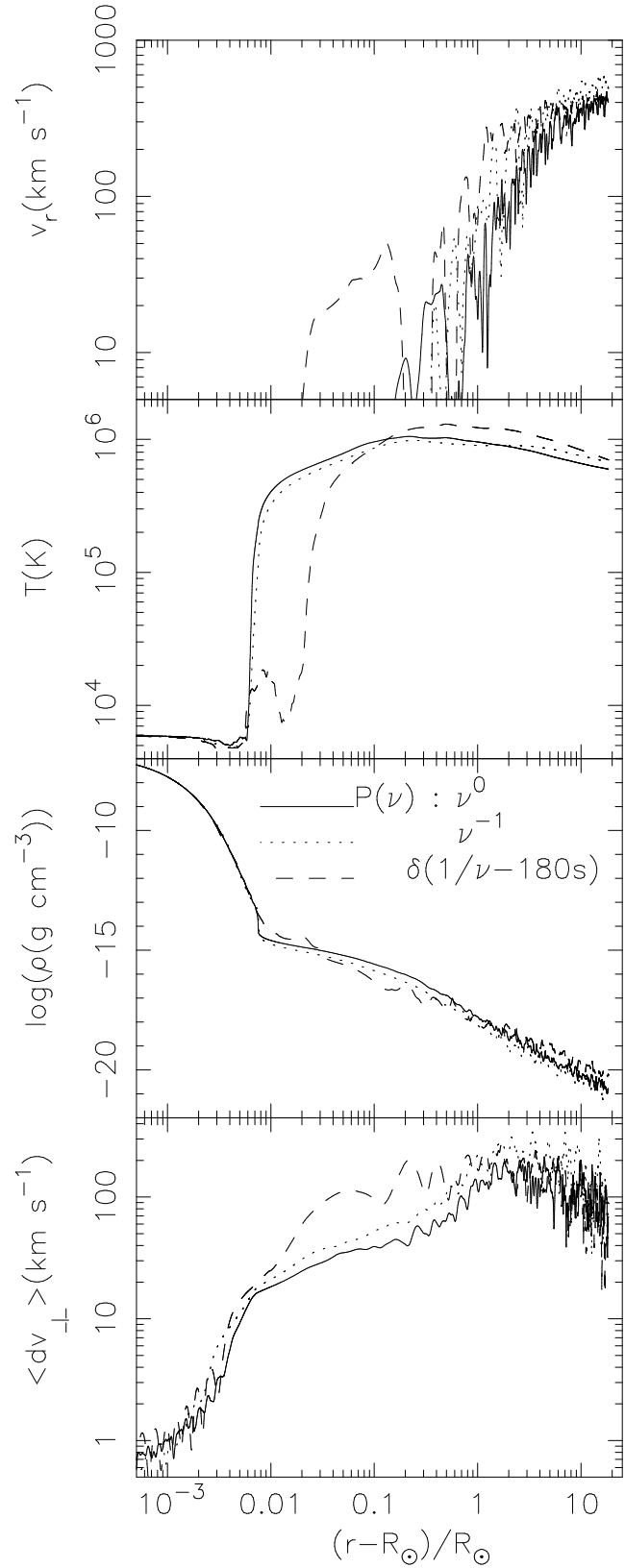}
\end{center}
\caption{Structures of corona and solar wind for various fluctuation 
spectrums at the photosphere. From top to bottom, we plot solar wind speed, 
$v_r$(km s$^{-1}$), temperature, $T$(K), density in logarithmic scale, 
$\log (\rho({\rm g\; cm^{-3}}))$, and rms transverse velocity, 
$\langle dv_{\perp}\rangle$(km s$^{-1}$). 
Dotted, solid, and dashed lines are results of $P(\nu)\propto 
\nu^{-1}$ (Run II), $\nu^{0}$ (Run III), and $\delta(1/\nu-180{\rm s})$ 
(Run IV), respectively. 
Each variable is averaged with respect to time during 28min (the longest 
wave peiriod considered).
}
\label{fig:dpwsp1}
\end{figure}

\begin{figure}
\figurenum{6} 
\begin{center}
\noindent\includegraphics[width=20pc]{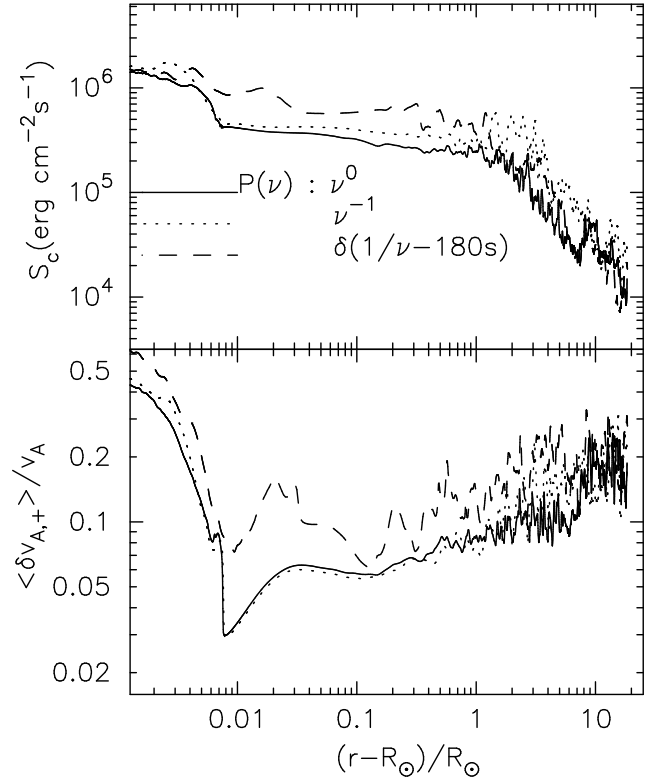}
\end{center}
\caption{Adiabatic constant, $S_c$(erg cm$^{-2}$s$^{-1}$), (top) and 
normalized amplitude, $\langle \delta v_{\rm A,+}\rangle /v_{\rm A}$, 
(bottom) of outgoing \Alfven waves for various 
spectrums. Dotted, solid, and dashed lines are results of $P(\nu)\propto 
\nu^{-1}$ (Run II), $\nu^{0}$ (Run III), and $\delta(1/\nu-180{\rm s})$ 
(Run IV), respectively. 
Each variable is averaged with respect to time during 28min (the longest 
wave peiriod considered).}
\label{fig:dpwsp2}
\end{figure}

We show dependences on the spectrum, $P(\nu)$, of the photospheric 
fluctuations. 
We here compare the results of 
$P(\nu) \propto \nu^{-1}$ ($6\times 10^{-4}<\nu < 2.5\times 10^{-2}$Hz; 
Run II), $\nu^{0}$(white noise; Run III), and purely
sinusoidal perturbation with a period of 180 s, $P(\nu) \propto 
\delta(\nu^{-1} -180{\rm s})$ (Run IV). 
Figure \ref{fig:dpwsp1} shows structure of the coronae and solar winds. 
The upper panel of Figure \ref{fig:dpwsp2} plots an adiabatic constant, 
$S_{\rm c}$, of the outgoing \Alfven wave derived from wave action 
\citep{jaq77}: 
\begin{equation}
S_c =
\rho \langle \delta v_{\rm A,+}^2 \rangle \frac{(v_r + v_{\rm A})^2}
{v_{\rm A}} \frac{r^2 f(r)}{r_c^2 f(r_c)}, 
\end{equation} 
where 
\begin{equation}
\delta v_{\rm A,+} = \frac{1}{2}\left(v_{\perp} - B_{\perp}/\sqrt{4\pi\rho}
\right)
\end{equation}
is amplitude of the outgoing \Alfven wave (Els\"{a}sser variables). 
The lower panel shows nonlinearity, $\langle \delta v_{\rm A,+}
\rangle /v_{\rm A}$ of the outgoing \Alfven waves. 

Figure \ref{fig:dpwsp1} shows that the plasma is heated up to the coronal 
temperature and accelerated to 400-600km/s at 0.1AU in all the Runs. 
Particularly, the results of $P(\nu)=\nu^{-1}$ and $\nu^{0}$ are 
quite similar. The attenuation of the outgoing \Alfven waves is due to the 
two kinds of the processes, the shocks and the reflection in our simulations. 
The shock dissipation per unit distance is larger for waves with 
larger $\nu$ because the number of shocks is proportional to wavenumber. 
On the other hand, \Alfven waves with smaller $\nu$ suffer the 
reflection more since the wavelengths are relatively larger compared to 
the variation scale of the \Alfven speed so that the wave shapes are 
deformed more. These two ingredients with respect to the $\nu$-dependence 
cancel in these two cases,  
and the variations of the energy and the amplitude become quite similar 
(Figure \ref{fig:dpwsp2}). The structures of the corona and the solar 
wind are also similar since they are consequences of the wave 
dissipation. 

The results of the purely sinusoidal fluctuation (Run IV) are slightly 
different from the other two cases. The TR is not so sharp as those in the 
other two Runs. 
The temperature firstly rises at $r-R_{\odot}=0.005R_{\odot}$, but 
drops slightly and again rises around  $r-R_{\odot}=0.02R_{\odot}$. 
Actually, the TR moves back and forth more compared to 
the other cases. This is because the shock heating takes place rather 
intermittently only at the wave crests of the monochromatic wave, 
while in the other cases the heating is distributed 
more uniformly by the contributions from waves with various $\nu$'s. 
As a result, the chromospheric evaporation occurs intermittently so that the 
TR becomes ``dynamical''. 
The sharpness of the TR is important in terms of the wave reflection. 
In the sinusoidal case (Run IV) the \Alfven waves becomes more transparent 
at the TR since the variation of the \Alfven speed is more 
gradual. Therefore, more wave energy transmits into the corona, avoiding  
the reflection at the TR (Figure \ref{fig:dpwsp2}). 
Accordingly, the temperature and density in the outer region become higher 
owing to the larger plasma heating (Figure \ref{fig:dpwsp1}).

\subsubsection{Wave Polarization}
\label{sec:wpl}
We examine the effect of the wave polarization. 
Linearly polarized \Alfven waves dissipate by the direct steepening 
\citep{hol82,suz04} to MHD fast shocks as well as by the parametric decay 
instability 
\citep{gol78}. 
On the other hand, circularly polarized components do not 
steepen, and it dissipates only by the 
parametric decay instability in our one-fluid MHD framework, 
so that the dissipation becomes less efficient. 

\begin{figure}
\figurenum{7} 
\begin{center}
\noindent\includegraphics[width=20pc]{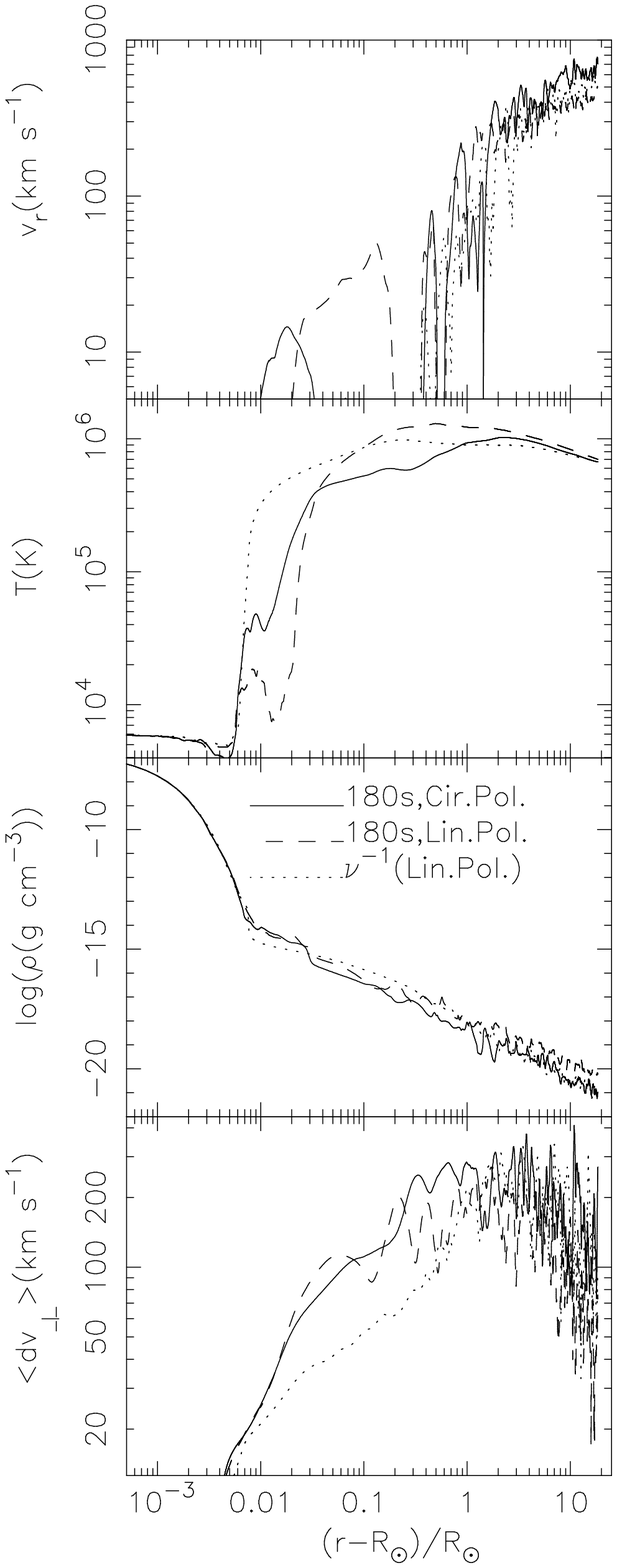}
\end{center}
\caption{Same as Figure \ref{fig:dpwsp1} but for different polarizations of 
the perturbations. Solid, dashed, and dotted lines are results of 
circularly polarized sinusoidal waves (Run V), linearly polarized sinusoidal 
waves (Run IV), and linearly polarized waves with power spectrum, 
$P(\nu)\propto \nu^{-1}$ (Run II), respectively. }
\label{fig:dpwpl1}
\end{figure}

\begin{figure}
\figurenum{8} 
\begin{center}
\noindent\includegraphics[width=20pc]{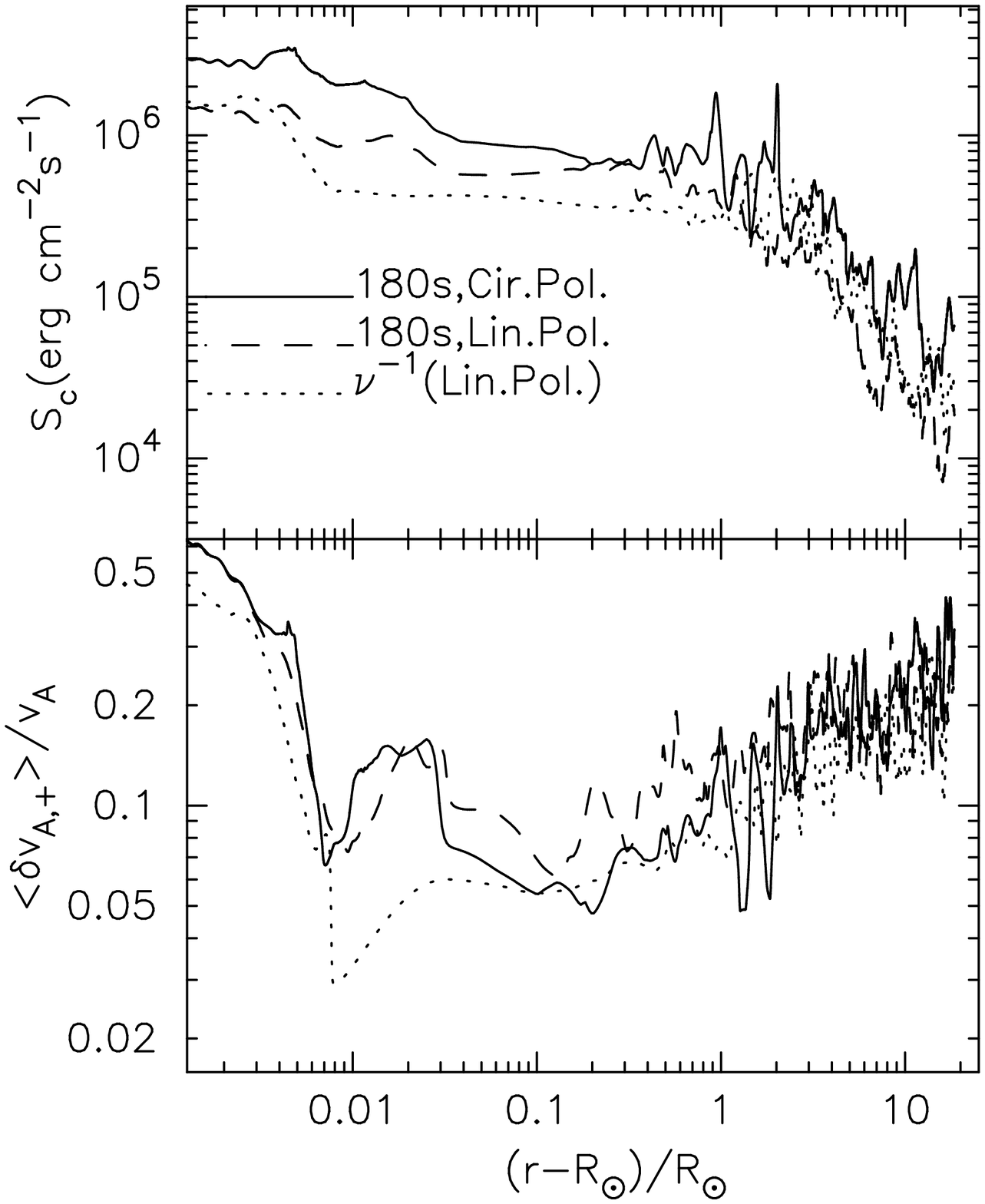}
\end{center}
\caption{Same as Figure \ref{fig:dpwsp2} but for different polarizations of 
the perturbations. Solid, dashed, and dotted lines are results of 
circular polarized sinusoidal waves (Run V), linearly polarized sinusoidal 
waves (Run IV), and linearly polarized waves with power spectrum, 
$P(\nu)\propto \nu^{-1}$ (Run II), respectively.  }
\label{fig:dpwpl2}
\end{figure}

Figures \ref{fig:dpwpl1} and \ref{fig:dpwpl2} compare the results of 
sinusoidal linearly polarized (Run IV) and circularly polarized 
(Run V) fluctuations with the same amplitude ($\langle dv_{\perp},0\rangle
=0.7$km s$^{-1}$). The results of Run II are also plotted for comparison. 
It is shown that the coronal heating and solar wind acceleration 
are still achieved by the dissipation of the 
circularly polarized \Alfven waves (Figure \ref{fig:dpwpl1}) although 
a large fraction of the wave energy remains at 
the outer boundary (Figure \ref{fig:dpwpl2}) owing to the less dissipative 
character as expected. 

A key ingredient for the efficient dissipation is the rapid decrease of the 
density owing to the gravity as discussed in \S\ref{sec:hawd}. 
The circularly polarized \Alfven waves 
which are initially sinusoidal are quickly deformed by the 
rapid variation of the density in the chromosphere and the low corona; 
simple sinusoidal waves cannot propagate any further unless the 
wavelengths are sufficiently short. 
The shape deformation indicates that the waves are partially reflected and 
the incoming \Alfven waves are generated.  
Compressive slow waves are also easily generated even without 
the steepening because the variation of the magnetic pressure is present 
owing to the large difference of the \Alfven speed within 
one wavelength \citep{bk96,od97,od98,gra02}.   
 
It has been generally believed that the decay instability of the \Alfven 
waves is not important in the context of the coronal heating and the solar 
wind acceleration. However, our simulation 
shows that this is not the case; the three-wave 
interactions are greatly 
enhanced by the long wavelength effect so that the sufficient wave 
dissipation occurs in the real solar corona.  

The case of the circularly polarized waves gives smaller coronal density 
and temperature since the wave dissipation 
is less effective. However, the solar wind speed is rather faster, up to 
700km s$^{-1}$ at 0.1AU, because the less dense plasma can be efficiently 
accelerated by transfer from the momentum flux of the \Alfven waves.

\subsubsection{Wave Amplitude}
\label{sec:dvdpn}
\begin{figure}
\figurenum{9} 
\begin{center}
\noindent\includegraphics[width=20pc]{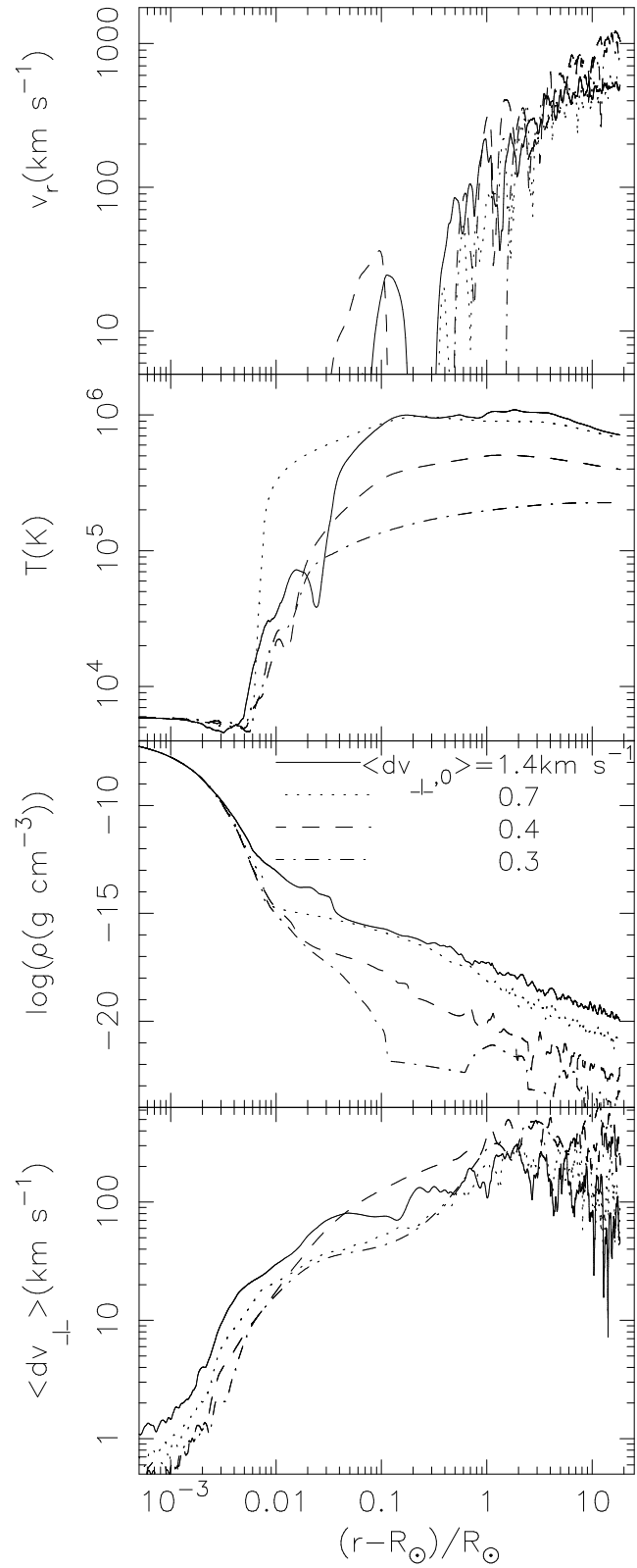}
\end{center}
\caption{Same as Figure \ref{fig:dpwsp1} but for different 
$\langle dv_{\perp,0} \rangle$. 
Solid, dotted, dashed, and dot-dashed lines are results of 
$\langle dv_{\perp,0}\rangle = 1.4$ (Run VI), 0.7 (Run II), 0.4 (Run VII), 
and 0.3(km s$^{-1}$) (Run IX), respectively. }
\label{fig:dpwam1}
\end{figure}

\begin{figure}
\figurenum{10} 
\begin{center}
\noindent\includegraphics[width=20pc]{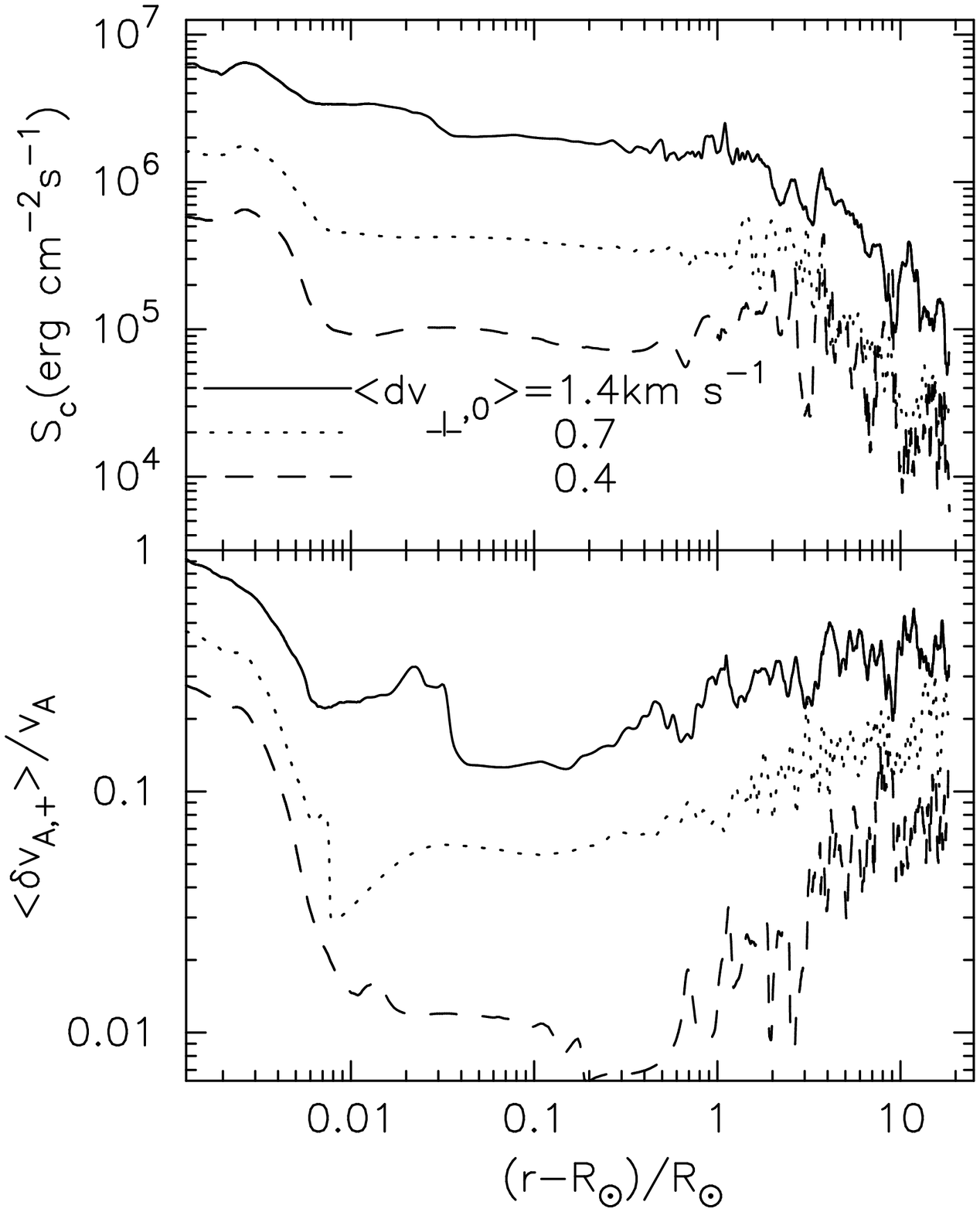}
\end{center}
\caption{Same as Figure \ref{fig:dpwsp2} but for different 
$\langle dv_{\perp,0}\rangle$. Solid, dotted, and dashed lines are results of 
$\langle dv_{\perp,0}\rangle = 1.4$ (Run VI), 0.7 (Run II), and 
0.4(km s$^{-1}$) (Run VII), respectively. }
\label{fig:dpwam2}
\end{figure}

We study dependences on the amplitudes of the input fluctuations 
at the photosphere. 
We compare the results of larger $\langle dv_{\perp,0} \rangle = 
1.4$km s$^{-1}$ (Run VI) and smaller $\langle dv_{\perp,0} 
\rangle = 0.4$km s$^{-1}$ (Run VII) cases with the fiducial case 
($\langle dv_{\perp,0} \rangle = 0.7$km s$^{-1}$; Run II) 
in Figures \ref{fig:dpwam1} and \ref{fig:dpwam2}.   
We also show the results of $\langle dv_{\perp,0} \rangle = 0.3$km s$^{-1}$ 
(Run IX) at $t=18.3$hr in Figure \ref{fig:dpwam1}, although  
quasi steady-state behavior is not achieved. 
At later time in Run IX, the density and temperature decrease with time 
as explained later. 

The maximum temperature of Run VII 
($\langle dv_{\perp,0} \rangle = 0.4$km s$^{-1}$) 
is $\simeq 5\times 10^5$K, which is cooler than the usual 
corona. The density is much lower than the fiducial
case by 1-2 orders of magnitude because the sufficient mass cannot supply 
into the corona by the chromospheric evaporation owing to the low 
temperature; the evaporation is drastically suppressed as $T$ decreases 
since the conductive flux ($F_c \propto T^{5/2}\frac{dT}{dr}$) sensitively 
depends on $T$. 
As a result, the mass flux ($\rho v_r$) becomes more than an order of 
magnitude lower than that of the present solar wind. 
These tendencies are more extreme in Run IX; 
The coronal temperature and density become further low in Run IX; 
$T\simeq 2\times 10^5$K and the density in the corona and solar wind is 
smaller by 3 orders of magnitude than the fiducial case.

This behavior can be understood by the wave dissipation 
(Figure \ref{fig:dpwam2}). The input wave energy ($\propto \langle 
dv_{\perp,0}^2 \rangle$) in Run VII is 1/3 of 
that of Run II.
However, the upper panel of Figure \ref{fig:dpwam2} shows that the remained 
$S_c$'s in both cases are similar at 0.1AU, which indicates that the \Alfven 
waves do not 
effectively dissipate in the smaller $\langle dv_{\perp,0} \rangle$ case. 
This is because the \Alfven speed ($=B/\sqrt{4\pi \rho}$) is larger 
due to the lower density and the nonlinearity, $\langle \delta v_{\rm A,+} 
\rangle /v_{\rm A}$, becomes weaker. 
As a result, the final energy flux of the solar wind at 0.1AU, mostly 
consisting of the kinetic energy ($\rho v_r \frac{v_r^2}{2}$), becomes 
one order of magnitude smaller although the ratio of the input wave 
energy is only 1/3. 

Once the coronal density starts to decrease, a positive feedback operates. 
Namely, the decrease of the density leads to weaker non-linearity of 
the \Alfven waves, which reduces the plasma heating by the wave dissipation. 
This decreases the coronal temperature, and further reduces the coronal 
density by the suppression of the chromospheric evaporation. 
This takes place in Run IX ($\langle dv_{\perp,0} \rangle=0.3$km s$^{-1}$) 
so that the coronal density and temperature continue to decrease at later 
time, instead of maintaining steady corona and solar wind.  

Our results show that $\langle dv_{\perp,0} \rangle\gtrsim0.4$km s$^{-1}$ is 
the criterion of the photospheric fluctuations for the formation of the stable 
hot plasma. To get the maximum coronal temperature $\gtrsim 10^6$K, 
$\langle dv_{\perp,0} \rangle\gtrsim 0.7$km s$^{-1}$ is required. 
Otherwise if $\langle dv_{\perp,0} \rangle\lesssim 0.3$km s$^{-1}$, the 
low-frequency \Alfven waves cannot maintain 
the hot corona, and the solar wind mass flux becomes drastically small; 
solar winds {\em virtually disappear} (see \S\ref{sec:dspsl}).

The solar wind speeds are slightly faster in the cases of smaller  
$\langle dv_{\perp,0} \rangle$ (accordingly smaller energy input),  
because the densities at the coronal bases are much  
smaller owing to the suppression of the chromospheric evaporation; 
the smaller amounts of the materials are accelerated to the faster speeds. 
The differences between the photospheric and coronal base densities are larger 
in the smaller $\langle dv_{\perp,0} \rangle$ cases and the amplification 
of $dv_{\perp}$ ($\propto \rho^{-1/4}$ if \Alfven waves are 
nondissipative in static media; e.g. Lamers \& Cassinelli 1999) 
is larger.  
As a result, $\langle dv_{\perp} \rangle$ at the coronal base 
is (weakly) anti-correlated with $\langle dv_{\perp,0} \rangle$ at the 
photosphere. Thus, our result is consistent with the previous calculations 
\citep{sl95,od98} that show the positively correlation between 
the \Alfven wave amplitude at the coronal base and the solar wind speed. 

Larger $\langle dv_{\perp,0} \rangle$ gives larger coronal density as shown 
in Run VI. 
The initial increase of the temperature starts from a deeper location around 
$r\simeq 0.005R_{\odot}$ than the other cases. Thanks to this,  
a decrease of the density is slower (larger pressure scale height) so that 
the density around $r=1.01R_{\odot}$ is two orders of magnitude larger than 
that of Run II. However, the temperature decreases slightly instead of a 
monotonical increase; it cannot go over the peak of the radiative cooling 
function at $T\simeq 10^5$K \citep{LM90} because the radiative loss is 
efficient owing to the large density. 
The second increase of the temperature begins 
from $r\simeq 1.03R_{\odot}$ and above there the corona is formed. 
The coronal density and temperature are larger than those in Run II. 
In particular 
the density in the outer region is 10 times larger, and the mass flux 
of the solar wind is larger by the same extent. 
The top panel of Figure \ref{fig:dpwam2} shows that the \Alfven waves 
dissipate more effectively in the larger $\langle dv_{\perp,0} \rangle$ case 
by the same mechanism as explained above for the smaller $\langle dv_{\perp,0}
\rangle$ cases. 

\begin{figure}
\figurenum{11} 
\begin{center}
\noindent\includegraphics[width=20pc]{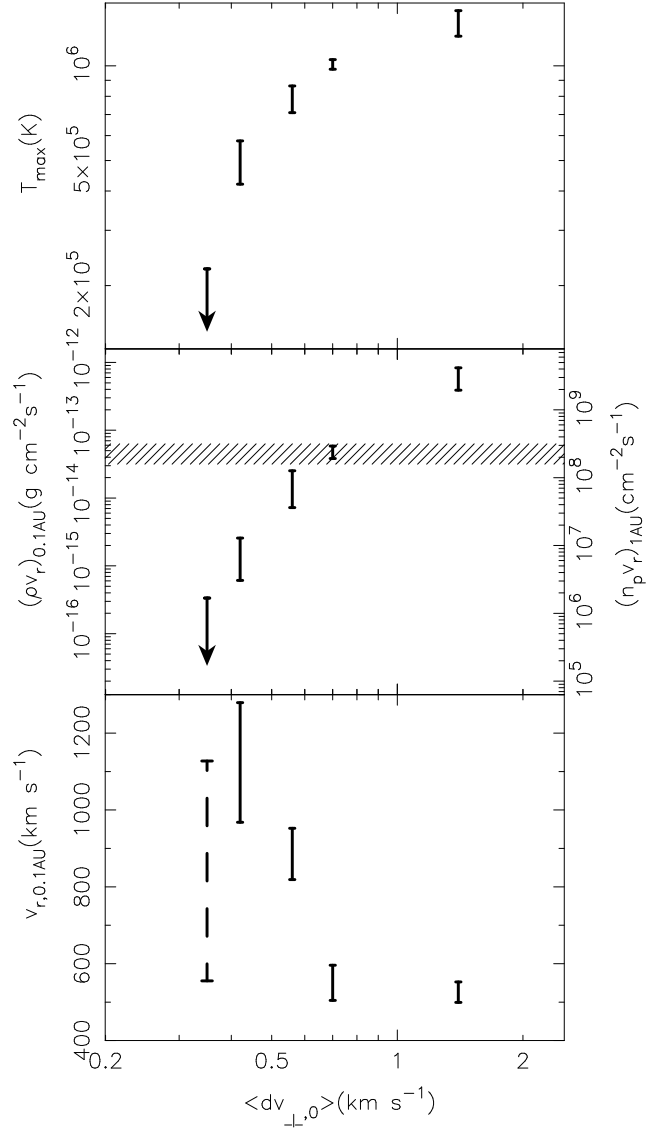}
\end{center}
\caption{The maximum temperature, $T_{\rm max}$(K), (top), the mass flux of 
solar wind at 0.1AU, $(\rho v_r)_{\rm 0.1AU}$(g cm$^{-2}$s$^{-1}$), (middle), 
and the solar wind speed at 0.1AU, $v_{\rm 0.1AU}$(km s$^{-1}$) (bottom). 
On the right axis 
of the middle panel, we show proton flux at 1AU, $(n_{\rm p}v_r)_{\rm 1AU}$
(cm$^{-2}$ s$^{-1}$) estimated by assuming the solar elemental abundance. 
The shaded region in the middle panel is observed proton flux at 1AU. 
The upper limits in the top and middle panels for $\langle dv_{\perp,0}\rangle
=0.3$km s$^{-1}$ indicates that stable corona with the sufficient mass 
supply cannot 
be maintained and both $T_{\rm max}$ and $(\rho v_r)_{\rm 0.1AU}$ decreases 
as the simulation proceeds. 
The wind speed also varies a lot, hence, the dashed error bar is used 
only for that case.    
}
\label{fig:dpwam3}
\end{figure}

Figure \ref{fig:dpwam3} summarizes the maximum coronal temperature, 
$T_{\rm max}$(K) (top), the solar wind mass flux at 0.1AU, 
$(\rho v_r)_{\rm 0.1AU}$(g cm$^{-2}$s$^{-1}$) (middle), and solar wind 
speed at 0.1AU, $v_{r,{\rm 0.1AU}}$(km s$^{-1}$), (bottom) 
as functions of $\langle dv_{\perp,0}\rangle$(km s$^{-1}$). 
On the right axis of the middle panel, we show the proton flux at 1AU, 
$(n_{\rm p}v_r)_{\rm 1AU}$(cm$^{-2}$s$^{-1}$). When deriving 
$(n_{\rm p}v_r)_{\rm 1AU}$ from $(\rho v_r)_{\rm 0.1AU}$, we use the 
relation of $\rho v_r \propto r^{-2}$, and assume the solar elemental 
abundance.
$T_{\rm max}$ and $(\rho v_r)_{\rm 0.1AU}$ have positive correlations with 
$\langle dv_{\perp,0} \rangle$.  
The dependence of $(\rho v_r)_{\rm 0.1AU}$ is quite steep because of the 
nonlinearity of the \Alfven waves as explained above. 
The dependence 
of $T_{\rm max}$ is more gradual due to the redistribution of 
the temperature by the thermal conduction. 
$v_{r,{\rm 0.1AU}}$ has weakly negative correlation with 
$\langle dv_{\perp,0} \rangle$ for $\langle dv_{\perp,0} 
\rangle\ge 0.4$(km s$^{-1}$) because in smaller 
$\langle dv_{\perp,0} \rangle$ cases the coronal density is further lower and 
smaller amounts of materials are accelerated more  
effectively. For $\langle dv_{\perp} \rangle = 0.3$(km s$^{-1}$), 
the coronal 
temperature and the mass flux of the solar wind becomes smaller later time and 
the wind speed varies quite a lot; we use upper limits for $T_{\rm max}$ 
and $(\rho v_r)_{\rm 0.1AU}$, and a dashed error bar for $v_{r,{\rm 0.1AU}}$. 
 
In our simulation we only consider the steady fluctuations with 
constant $\langle dv_{\perp,0} \rangle$'s to study the basic relation between 
the footpoint fluctuations and the properties of the solar winds. 
In the real 
situation, however, $\langle dv_{\perp,0} \rangle$ would vary in time. 
Our results imply that a small temporal variation of $\langle dv_{\perp,0} 
\rangle$ at the surface possibly leads to large time dependent 
behavior of the solar wind. As one example we would like to discuss 
an event of solar wind disappearance in the following section. 

\subsubsection{Disappearance of Solar Wind}
\label{sec:dspsl}
On May 10-12, 1999, observed solar wind density near 
the earth drastically decreases well below 1 cm$^{-3}$ for $\sim$ a day and to 
0.1 cm$^{-3}$ at the lowest level, in comparison with 
the typical value $\sim 5$ cm$^{-3}$  \citep{lrp00, smt01}.  
Various mechanisms are widely discussed to explain this disappearance event
\citep{usa00,ric00,cro00}. 
In \S \ref{sec:dvdpn}, 
we have shown that the solar wind density sensitively 
depends on the amplitude of the photospheric fluctuation provided that 
the nonlinear dissipation of low-frequency \Alfven waves operates in 
the solar wind heating and acceleration.  According to Figure \ref{fig:dpwam3}
the solar wind density and mass flux becomes $\sim 100$ and $\sim 50$ times 
smaller respectively if $\langle dv_{\perp,0}\rangle$ changes from 
0.7km s$^{-1}$ to 0.4km s$^{-1}$; a small variance of the energy injection 
at the solar surface leads to a large variation of the solar wind density 
because of the nonlinearity.  
Therefore, we can infer that if the amplitude of the photospheric turbulence 
becomes $\sim$ 1/2 during $\sim$a day, this event of the solar wind 
disappearance is possibly realized.

The observation shows that the velocity of the sparse plasma discussed 
above is lower than that of the surrounding plasma, which is inconsistent 
with our results (\S \ref{sec:dvdpn}). 
We are speculating that the stream interaction would resolve the 
inconsistency. The sparse region is easily blocked by the preceding dense 
plasma in the spiral magnetic fields (Parker spiral) even if its 
speed in the inner region (say $\sim 0.1$AU) is faster \citep{usa00}, 
because the ram pressure is much smaller than that of the surrounding plasma. 
For quantitative arguments, however, two dimensional modeling is required.    

\subsection{Flux Tube properties}
\label{sec:flxtp}
It is widely believed that field strength and geometry of open flux 
tubes are important parameters that control  
the solar wind speed. \citet{ws90,ws91} showed that the solar wind speed at 
$\sim$ 1AU is anti-correlated with $f_{\rm max}$ from their long-term 
observations as well as by a simple theoretical model. \citet{od98} 
showed this tendency by time-dependent simulations as well.   
\citet{fsz99} claimed that the wind speed should be positively correlated 
with $B_{r,0}$ by a simple energetics consideration. 
\citet{kfh05} 
have found that the solar wind velocity 
is better correlated with the combination of these two parameters, 
$B_{r,0}/f_{\rm max}$, than 
$1/f_{\rm max}$ or $B_{r,0}$ from the comparison of the outflow speed 
obtained by their interplanetary 
scintillation measurements with observed photospheric field strength. 
\citet{suz04} and \citet{suz06} also pointed out that $B_{r,0}/f_{\rm max}$ 
should be the best control parameter provided that the \Alfven waves play 
a dominate role in the coronal heating and the solar wind acceleration. 

\begin{figure}
\figurenum{12} 
\begin{center}
\noindent\includegraphics[width=20pc]{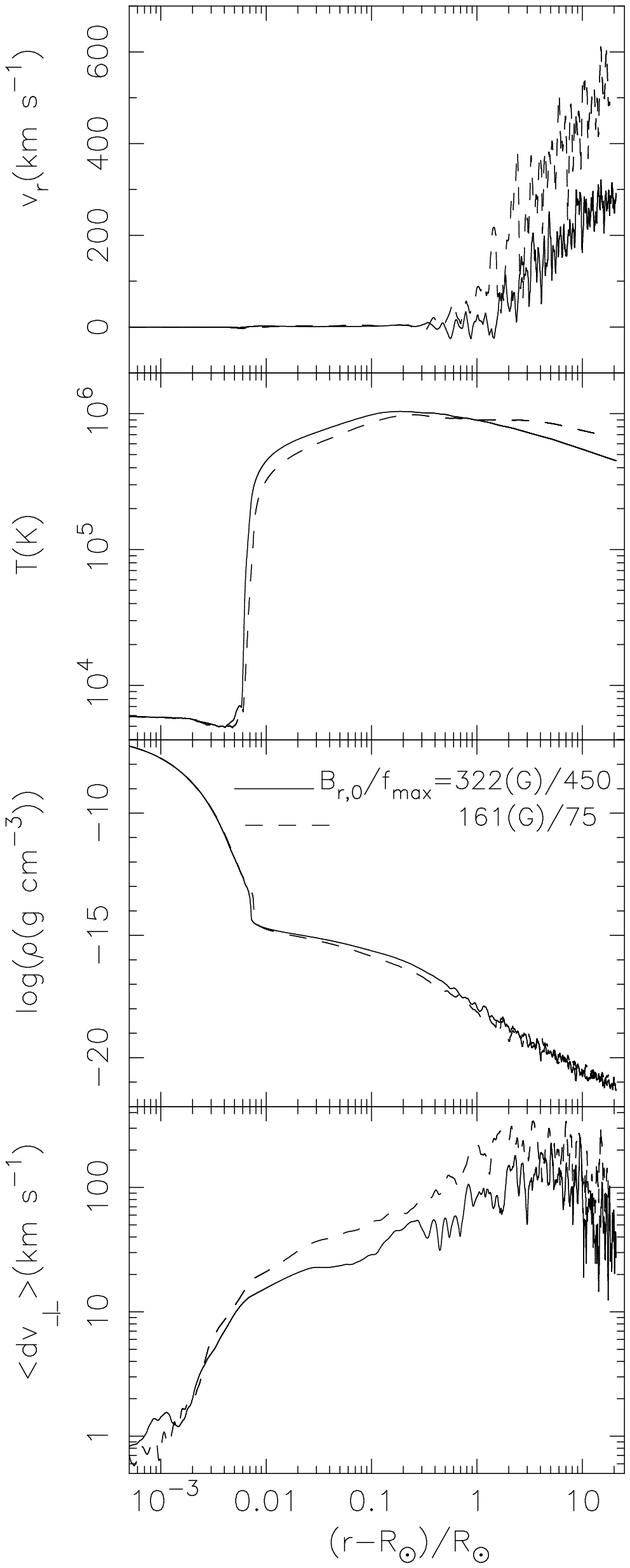}
\end{center}
\caption{Same as Figure \ref{fig:dpwsp1} but for different $B_{r,0}/
f_{\rm max}$. Solid and dashed lines are results of $B_{r,0}({\rm G})/
f_{\rm max} = 322/450$ (Run VIII) and 161/75 (Run II), respectively.}
\label{fig:dpbf1}
\end{figure}

\begin{figure}
\figurenum{13} 
\begin{center}
\noindent\includegraphics[width=20pc]{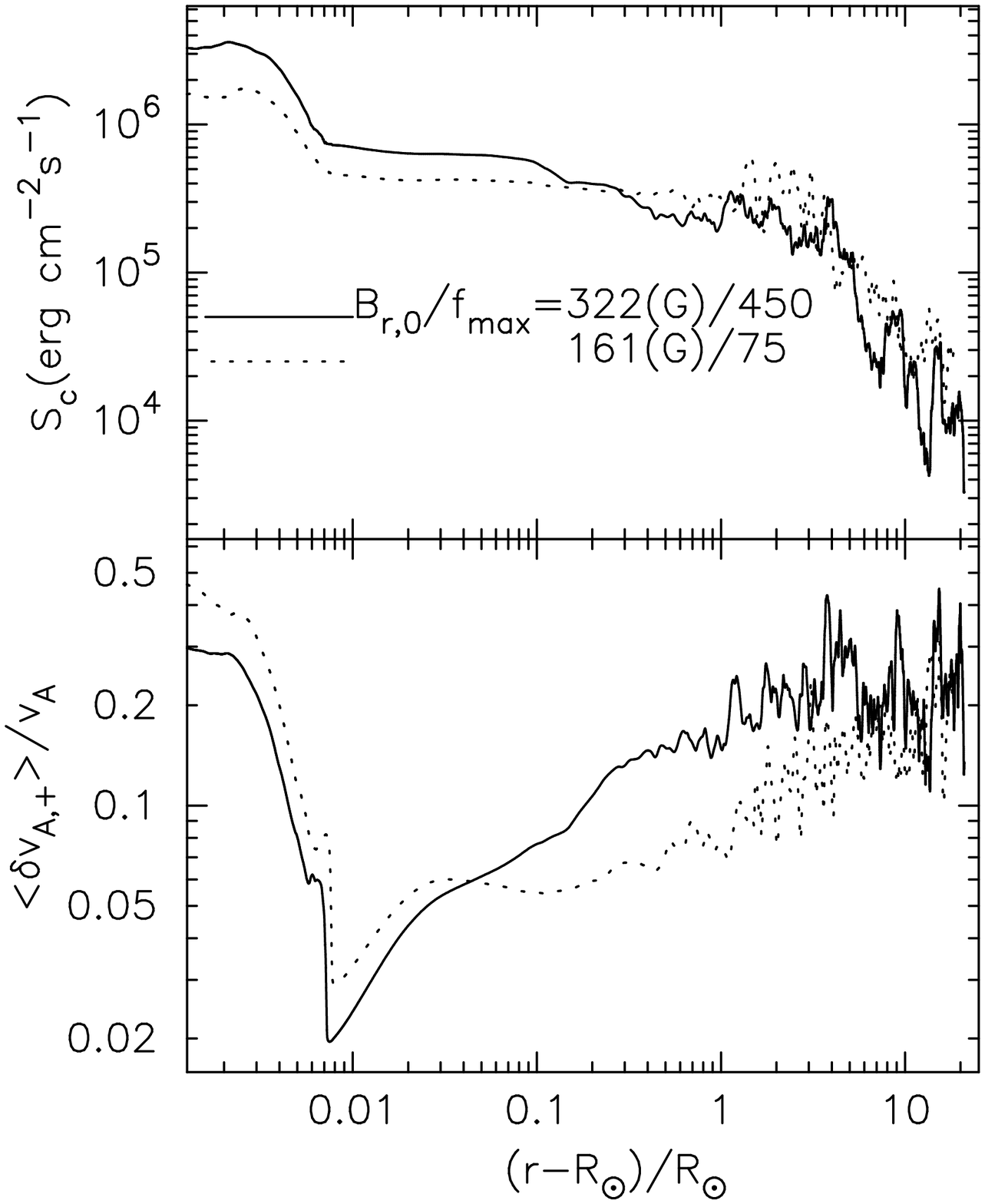}
\end{center}
\caption{Same as Figure \ref{fig:dpwsp2} but for different $B_{r,0}/
f_{\rm max}$. Solid and dashed lines are results of $B_{r,0}({\rm G})/
f_{\rm max} = 322/450$ (Run VIII) and 161/75 (Run II), respectively.}
\label{fig:dpbf2}
\end{figure}

Here we test the last scenario by our simulation. 
In Figures \ref{fig:dpbf1} and \ref{fig:dpbf2}, we compare the results of 
a flux tube with smaller $B_{r,0}/f_{\rm max} 
=322{\rm (G)}/450$ (Run VIII) with the fiducial case ($B_{r,0}/f_{\rm max} 
=161{\rm (G)}/75$) (Run II) for the polar coronal hole. 
We chose the parameters of Run VIII with equatorial and mid-latitude 
coronal holes in mind; they are surrounded by closed 
structures, and thus, have moderately larger photospheric field 
strength and much larger flux tube divergence than the polar coronal hole. 

Figure \ref{fig:dpbf1} shows that Run VIII gives much slower solar wind 
speed which is consistent with \citet{kfh05}, \citet{suz04}, and 
\citet{suz06}, with slightly hotter and denser corona. 
These results can be understood by positions of 
the wave dissipation. The upper panel of Figure \ref{fig:dpbf2} indicates that 
the outgoing \Alfven waves dissipate more rapidly in the smaller 
$B_{r,0}/f_{\rm max}$ case. 
This is because the nonlinearity of the \Alfven waves, $\langle \delta 
v_{\rm A,+}\rangle /v_{\rm A}$
is larger due to smaller $v_{\rm A} \propto B_r$($\propto 
B_{r,0}/f_{\rm max}$ in
the outer region where the flux tube is already super-radially open) 
even though the absolute amplitude ($\langle d v_{\perp}\rangle $) is smaller 
(
bottom panel of Figure \ref{fig:dpbf1}).   
As a result, more wave energy dissipate in the smaller $B_{r,0}/f_{\rm max}$ 
case in the subsonic region and less energy remains in the supersonic region. 
In general, energy and momentum inputs in the supersonic region gives higher 
wind speed, while those in the subsonic region raises the mass flux 
($\rho v_r$) of the wind by an increase of the density \citep{lc99}.  
Therefore, the smaller $B_{r,0}/f_{\rm max}$ case gives slower wind with 
higher coronal density, whereas the solar wind density in the outer region 
is similar to that of the larger $B_{r,0}/f_{\rm max}$ case on account of 
the dilution of the plasma in the more rapidly expanding flux tube.   
We can conclude that $B_{r,0}/f_{\rm max}$ controls the solar wind speed 
and the coronal density through the nonlinear dissipation of the 
\Alfven waves.  

\subsection{Summary of Parameter Studies}
\label{sec:smps}
We have examined the dependences of the coronal and wind properties on 
the various wave and flux tube parameters in \S \ref{sec:viw} 
and \S \ref{sec:flxtp}. 
One of the important results is that we do not find any subsonic 'breeze' 
solution in our simulations. All the Runs with $\langle dv_{\perp,0}\rangle 
\gtrsim 0.3$km s$^{-1}$ show the transonic feature, 
hence, transonic solar winds are  
natural consequences of the dissipation of the low-frequency \Alfven waves.  
Smaller energy injection only reduces the density (and consequently 
the mass flux) of the wind. The wind speed itself is rather slightly  
faster for smaller wave energy because the coronal base density becomes 
lower owing to the suppression of the chromospheric evaporation.  
 
From \S \ref{sec:wsp} and \S \ref{sec:wpl}, we can conclude that the 
dependences on the spectra and polarizations of the input fluctuations are 
weak as long as the low-frequency ($\lesssim 0.05$Hz) waves are considered. 
$\langle dv_{\perp,0}\rangle$ and $B_{r,0}/f_{\rm max}$ are the important 
control parameters, and the dependences are summarized below: 
\begin{itemize}
\item{The coronal temperature and the density in the corona and solar wind 
are mainly determined by the photospheric amplitude, 
$\langle dv_{\perp,0}\rangle$. 
The corona with temperature $\gtrsim 10^6$K is formed if $\langle dv_{\perp,0} 
\rangle \gtrsim 0.7$km s$^{-1}$. Larger $\langle dv_{\perp,0}\rangle$ gives 
higher density in the corona and the solar wind. 
If $\langle dv_{\perp,0}\rangle\lesssim 0.3$km s$^{-1}$ 
the hot plasma cannot be maintained and the mass flux 
of the solar wind is unrealistically small.}
\item{The solar wind speed is mainly controlled by $B_{r,0}/f_{\rm max}$; 
faster winds come from open flux tubes with larger $B_{r,0}/f_{\rm max}$.}
\end{itemize}

\section{Fast and Slow Solar Winds}

The observed solar winds can be categorized into two distinctive 
types. 
One is the fast solar wind which is mainly from polar coronal holes 
and the other is the slow wind which is from mid- to low-latitude regions. 
Apart from the velocity difference, both density of solar winds 
\citep{phi95} and freezing-in temperatures of heavy ions \citep{gei95} 
which reflect the coronal temperatures are anti-correlated with velocities. 
Origins of the slow solar wind are still in debate;  
there are mainly two types argued (see Wang et al.2000 for review). 
One is the acceleration due to intermittent break-ups of the 
cusp-shaped closed fields in the equatorial region (e.g. Endeve, Leer, \& 
Holtzer 2003). The other is the acceleration in open flux tubes with large 
areal expansions in low- and mid-latitude regions.
\citet{kfo99} detected 
low-speed winds with single magnetic polarities, originating from 
open structure regions located near active regions, which indicates 
a certain fraction of the slow stream is coming from open regions, 
similarly to the fast wind. 
In this paper we focus on the latter case with respect to the slow 
solar wind. 


Based on our parameter studies summarized in \S \ref{sec:smps}, 
we can infer the suitable 
$\langle dv_{\perp,0}\rangle$ and $B_{r,0}/f_{\rm max}$ for the slow solar 
wind by comparing with those for the fast wind (Run I/II; also in paper I); 
the slow wind requires larger 
$\langle dv_{\perp,0}\rangle$ and smaller $B_{r,0}/f_{\rm max}$ than the fast 
wind. The adopted parameters are summarized in table 1 (Run X).  

\begin{figure}
\figurenum{14} 
\begin{center}
\noindent\includegraphics[width=20pc]{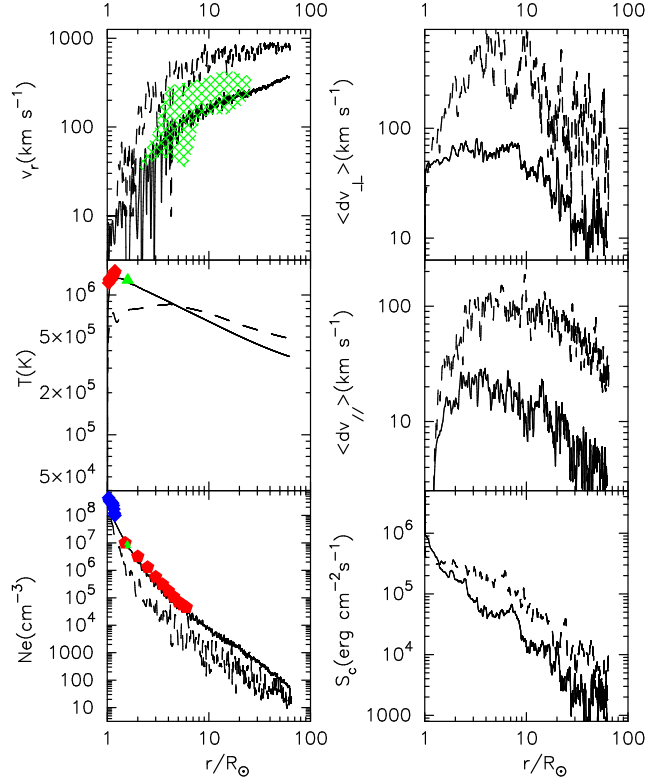}
\end{center}
\caption{Simulation results of the slow (Run X; solid lines) and 
fast (Run I; dashed lines) solar wind models in comparison with slow-wind 
observations. 
On the left from top to bottom, outflow velocity, $v_r$(km s$^{-1}$), 
temperature, $T$(K), and electron density, $N_e$(cm$^{-3}$), are 
plotted with observation of mid- to low-latitude regions where the slow 
wind comes from. (see Figure \ref{fig:obscmp} for the fast-wind 
observations).   
On the right from top to bottom rms transverse velocity, $\langle dv_{\perp} 
\rangle$(km s$^{-1}$), rms longitudinal velocity, $\langle dv_{\parallel} 
\rangle$(km s$^{-1}$), and the adiabatic constant, 
$S_c$(erg cm$^{-2}$s$^{-1}$) of the outgoing \Alfven waves are plotted. 
{\it Observational data; top-left} : Shaded region is 
observational data in the streamer belt \citep{she97}.  
{\it Middle-left} : Filled diamonds and triangles are electron temperature 
obtained from the line ratio of Fe XIII/X in the mid-latitude streamer by 
CDS/SOHO and UVCS/SOHO respectively \citep{pbp00}
{\it bottom-left} : Filled diamonds and triangles 
are data respectively from CDS and UVCS on SOHO observation of the 
mid-latitude streamer \citep{pbp00}, 
and filled pentagons derived from observation of the total brightness in the 
equator region by LASCO/SOHO \citep{hvh01}. }
\label{fig:fwsw}
\end{figure}

In the panels on the left side of Figure \ref{fig:fwsw} we compare the results 
of Run X (slow wind) and Run I (fast wind) overlayed with recent 
observations of slow solar winds (see Figure \ref{fig:obscmp} for the 
observations of fast winds). 
The temperature and density of the slow wind case becomes larger on account of 
the larger $\langle dv_{\perp,0}\rangle$, which explains the observations. 
On the other hand, smaller $B_0/f_{\rm max}$ gives slower terminal speed. 
As a result, the observed anti-correlation of the wind speed and the coronal 
temperature \citep{sm03} is well-explained by our simulations.   
In the slow wind case (Run X), the acceleration of the outflow is more 
gradual, and it is not negligible in $r \gtrsim 20_{\odot}$ 
(e.g. Nakagawa et al.2005). 

In the panels on the right of Figure \ref{fig:fwsw}, we show the properties 
of the wave dissipation. 
The top right panel compares transverse amplitude, 
$\langle dv_{\perp} \rangle$, averaged over 28 minutes in the fast and slow 
wind cases. 
$\langle dv_{\perp} \rangle$ is larger in  
the fast wind. This is because the \Alfven waves become less dissipative in 
the fast wind conditions; larger $B (\propto B_0/f_{\rm max})$ and 
smaller $\rho$ give larger $v_{\rm A}$ so that the nonlinearity of the 
\Alfven waves, $\langle \delta v_{\rm A,+}\rangle /v_{\rm A}$, is 
systematically smaller. 
As a result, the \Alfven wave dissipates more slowly in the fast wind case, 
which is shown in the plot of $S_c$ of the wave (the bottom right panel of 
Figure \ref{fig:fwsw}).  
This tendency of the larger $\langle dv_{\perp} \rangle$ in the fast wind 
is expected to continue to the outer region, $\gtrsim 1{\rm AU}$. 
Various in situ observations in the solar wind plasma 
near the earth also show that the fast wind contains more \Alfvenic 
wave components (see Tsurutani \& Ho 1999 for review), which can be 
explained via the nonlinear dissipation of \Alfven waves based on 
our simulations. 

The middle right panel of Figure \ref{fig:fwsw} shows longitudinal 
fluctuation, $\langle dv_{\parallel} \rangle$, averaged over 28 minutes. 
$\langle dv_{\parallel} \rangle$ reflects the amplitude 
of slow MHD waves generated from the \Alfven waves via the nonlinear 
process (\S \ref{sec:hawd}). 
Slow waves are identified in polar regions \citep{ofm99} and 
low-latitude regions \citep{sak02} at low altitudes 
($r\lesssim 1.2R_{\odot}$).  The observed amplitudes are still small 
($\simeq 7.5$km s$^{-1}$ in the polar regions and $\simeq 0.3$km $^{-1}$ 
in the low-latitude regions) because of the large density there, which 
is consistent with our results.  
What is more interesting is longitudinal fluctuations in outer 
regions. our results exhibits that $\langle dv_{\parallel} 
\rangle$ in the solar wind plasma is not small. 
Typically, the simulations give 
$\langle dv_{\parallel}\rangle \sim 100$km/s in $3R_{\odot} \lesssim r 
\lesssim 10R_{\odot}$ of the fast wind and 
$\langle dv_{\parallel}\rangle \sim 20$km/s in $2R_{\odot} \lesssim r 
\lesssim 10R_{\odot}$ of the slow wind, 
whereas these might be modified if we take into account 
multi-dimensional effects (\S\ref{sec:lmt}).  
This is directly testable by in situ measurements of future missions, 
Solar Orbiter and Solar Probe, 
which will approach to $\sim$45 and 4 $R_{\odot}$, respectively, 
corresponding to the inside of our computation domain.  

Our simulations show that the different types of the solar winds can be 
explained by the single process, the dissipation of the low-frequency \Alfven 
waves, although we do not intend to exclude other possibilities.  
The varieties of the solar winds are due to the varieties of the footpoint 
amplitudes as well as the magnetic fields and geometry of the flux tubes.   
Our choice of $B_{r,0}/f_{\rm max}$ is consistent with 
the obtained data by \citet{kfh05} 
who report that the slow winds 
are mainly from the low-to mid-latitude coronal holes with smaller 
$B_{r,0}/f_{\rm max}$. In contrast, quantitative arguments on 
$\langle dv_{\perp,0} \rangle$ need more detailed fine scale observations 
of the 
magnetic fields with high cadence by future telescopes such as Solar-B.  


\section{Discussions}
\label{sec:dis}

\subsection{Wave Generation}
In this paper we have assumed that the origin of the waves is the steady 
turbulent motions at the photosphere. However, transient activities 
are also expected to play a role in the wave generation. 
\citet{str99} proposed small flare-like events triggered by magnetic 
reconnections of closed loops in the chromosphere excite MHD waves 
in the corona. 
\citet{miy05} also show that interactions between open field lines and 
emergent flux tubes excite \Alfven waves at a location above the photosphere. 
These waves could directly heat up the corona since 
they do not suffer the attenuation in the chromosphere and the TR. 

An important issue for the wave generation by these transient events 
is the energetics. The energy release from each event is generally thought 
to be small, categorized as a micro- or nano-flare \citep{pj00,kt01}. 
To clarify 
how these waves dominantly work in the heating and acceleration of the 
solar wind, we should determine the frequency of these events. 
These small-scale events might be important in the 
heating of not only the open coronal holes but the closed regions. 
The determination of the total energy release from small transient events 
by future observations is quite important in terms of the coronal heating 
in various portions of the corona. 

\subsection{Limitation of Our Simulations} 
\label{sec:lmt}
We have shown by the self-consistent simulations that the dissipation of the 
low-frequency \Alfven waves through the generation of the 
compressive waves and shocks is one of the solutions for the heating and 
acceleration of the plasma in the coronal holes.    
However, the validity of the 1D MHD approximation 
needs to be examined.

We think that the MHD approximation is appropriate as a whole 
for studies of the average properties of the plasma. 
Let us estimate the Larmor radius, $l_{\rm Larmor}$, of the protons which 
compose the main part of the plasma due to 
fluctuations of field lines, $B_{\perp}$, under the coronal condition. 
Typically, $B_\perp \sim 0.1 B_r \sim 0.1$G, and then  
$l_{\rm Larmor} \sim 0.1$km for a thermal proton with $\sim 100$km s$^{-1}$. 
On the other hand the shortest wavelength, $\lambda_{\rm min}$, of the 
\Alfven waves we are considering is 
$\lambda_{\rm min}\sim 3\times 10^4$km (a period of 20s and 
$v_{\rm A}\sim 1500$km s$^{-1}$.) and 
the simulation grid size in the corona is $\sim 3\times 10^3$km ($\sim 1/10$ 
of $\lambda_{\rm min}$); the thermal 
protons are supposed to be well randomized through turbulent magnetic fields 
on the scales we are dealing with. 

In our simulations the heating of the plasma is done by MHD shocks 
(\S \ref{sec:stup}). 
Our simulations cannot resolve detailed structures of shocks since the size of 
the grid is larger than the width of the shock front which is an order of 
a Larmor radius.  
However, the global structure of shocks can be appropriately treated by our 
simulations.   
The thermal particles could be `trapped' around the shock regions by the 
random magnetic fields so that the global properties of shocks would satisfy 
the MHD condition. 
Therefore, the shock heating rate calculated 
in our MHD simulations is supposed to give a reasonable estimate in the 
coronal region. In the outer region ($r\gtrsim 2R_{\odot}$), $l_{\rm 
Larmor}$ ($\propto 1/B$) increases and $\lambda_{\rm min}$ 
($\sim v_{\rm A}/\nu_{\rm max} \propto v_{\rm A} 
\propto B/\sqrt{\rho}$) decreases as $r$ increases. Even around $0.3$AU, 
$l_{\rm Larmor}$ is about $\sim$1/10 of $\lambda_{\rm min}$; 
the approximation still remains reasonable.    

However, we should be cautious about the fact that the shocks are 
collisionless; 
mean free path due to Coulomb collisions, which is $\sim 100$km for 
electron-electron collisions and even larger for electron-proton and 
proton-proton collisions under the typical coronal condition 
($T=10^6$K and $n=10^9$cm$^{-3}$), is much larger than $l_{\rm Larmor}$ 
so that the particles are not thermalized 
by Coulomb collisions but only randomized by fluctuating fields.   
In such a condition, the particle distribution function possibly deviates from 
the Maxwell-Boltzmann distribution, which is actually observed in solar wind 
plasma \citep{mar82}. 
Non-thermal, or even supra-thermal, particles might modify 
the heating rate, if their amount is not 
negligible. The shock energy is transfered not only to the thermal heating 
but also to acceleration of the nonthermal component. 
Although this is beyond the scope of the present paper, this issue should be 
carefully considered in more detailed models.

Effects of multi-component plasma also need to be taken into account 
\citep{lsv01,ofm04}.
The thermal conduction and radiative cooling are mainly involved with 
electrons, and the electron temperature becomes systematically lower than 
the ion temperature without effective thermal coupling by Coulomb collisions. 
The observed proton temperature is actually higher than the electron 
temperature even around a few solar radii \citep{ess99}. 
The temperature in our simulations represents the electron temperature because 
it is mainly determined by the thermal conduction and radiative cooling. 
The actual thermal pressure 
is supposed to be higher than the simulated temperature due to 
the higher proton temperature. Therefore, the acceleration profile might be 
slightly modified whereas the the dynamics are mostly controlled by the 
magnetic pressure of the \Alfven waves.   

Collisionless processes would also be important 
for the wave dissipation \citep{tsu05}.  
The heating rates by the dissipation of \Alfven waves are different for  
electrons and ions \citep{hl75,tss05}. 
The compressive waves generated from the \Alfven waves would suffer 
transit-time damping by interactions between magnetic mirrors and surfing 
particles \citep{bar66,suz06}.    
If high-frequency (ioncyclotron) waves are produced by frequency cascade, 
the heating sensitively depends on mass-to-charge ratios of particles 
\citep{dh81,mgr82}. 
It is believed that this process can explain the observed 
preferential heating of the perpendicular temperature of the heavy ions 
\citep{kol98}, while there are potential difficulties in the proton 
heating as discussed in \S\ref {sec:int}.

In the chromosphere, the gas is not fully ionized. In such circumstances, 
the friction due to ion-neutral collisions might be important in 
the dissipation of \Alfven waves \citep{lvc04,yl04}. 

Generally, the shock dissipation tends to be overestimated in 1D simulations 
because the waves cannot be diluted by the geometrical expansion.  
On the other hand, there are other mechanisms of the wave damping due to the 
multi-dimensionality \citep{ofm04}, such as 
phase mixing \citep{hp83,nrm98,dem00} and refraction \citep{bog03}. 
Moreover, the solar wind plasma is more or less turbulent and it must be 
important to take into account the plasma heating through cascades of 
\Alfvenic turbulences in the transverse direction \citep{gs95,mat99,oug01}. 
Studies of the imbalanced cascade, in which a wave component in one direction 
has larger energy than that in the counter direction, seem to be important for 
us to understand transport phenomena in the solar wind plasma 
\citep{lg03}, though it has not been fully understood yet. 

Therefore, the \Alfven waves might also be dissipated by mechanisms different 
from those included in our simulations.
Accordingly, variation of wave amplitudes might be modified when these 
additional dissipation processes are considered.    
Self-consistent simulations including these various processes 
remain to be done in order to arrive at final conclusion.

\section{Summary}
We have performed parametric studies 
of the coronal heating and 
solar wind acceleration by the 
low-frequency \Alfven waves in open coronal holes. 
We have performed 1D MHD numerical simulations from the 
photosphere to 0.3 or 0.1AU. The low-frequency \Alfven waves with 
various spectra, polarization, and amplitude are 
generated by the footpoint fluctuations of the magnetic field lines.  
We have treated the wave propagation and dissipation, and the heating and 
acceleration of the plasma in a self-consistent manner. 

We have found that the transonic solar winds are accomplished in all the 
simulation Runs. Smaller energy injection does not reduce the 
outflow speed but the density to maintain the transonic feature. 
The atmosphere is heated up to 
the coronal temperature ($\gtrsim 10^6$K) if the photospheric amplitude, 
$\langle dv_{\perp,0} \rangle \gtrsim 0.7$km s$^{-1}$. 
Otherwise, the temperature and density become much lower than the present 
coronal values. 
If $\langle dv_{\perp,0} \rangle \lesssim 0.3$km s$^{-1}$, 
the temperature becomes 
less than a few $10^5$K and the sufficient mass cannot be supplied into 
the corona owing to the suppression of the chromospheric evaporation. 
The stable hot corona cannot be maintained any longer, and  
the mass flux of the solar wind becomes at least 3 orders of magnitude 
smaller than the observed value of the present solar wind. 
This shows that the solar wind almost 
disappears only by reducing the photospheric fluctuation amplitude by half.  
On the other hand, the case with $\langle dv_{\perp,0} 
\rangle = 1.4$km s$^{-1}$  gives 10 times larger density than the fiducial 
case ($\langle dv_{\perp,0} \rangle = 0.7$km s$^{-1}$ ). 
These sensitive behaviors of the solar wind mass flux on 
$\langle dv_{\perp,0} \rangle$ can be explained via the nonlinear dissipation 
of \Alfven waves. 
Our simulations have also confirmed that the positive correlation of 
the solar wind speed with $B_{r,0}/f_{\rm max}$ obtained by 
\citet{kfh05}. 

We have finally pointed out that both fast and slow solar winds can 
be explained even solely by the dissipation of the low-frequency \Alfven waves 
with different $\langle dv_{\perp,0}\rangle$ and $B_{r,0}/f_{\rm max}$, 
whereas we do not intend to exclude other possibilities. 
Fast winds are from flux tubes with larger $B_{r,0}/f_{\rm max}$ and 
smaller $\langle dv_{\perp,0}\rangle$, while slow winds are from flux 
tubes with smaller $B_{r,0}/f_{\rm max}$ and larger $\langle dv_{\perp,0}
\rangle$. These choices naturally explain the following observed tendencies: 
(i) the anti-correlation of the 
solar wind speed and the coronal temperature, and (ii) the larger amplitude 
of \Alfven waves in the fast wind in the interplanetary space. 
The tendency with respect to $B_{r,0}/f_{\rm max}$ is consistent 
with the observed trend \citep{kfh05}. To determine $\langle dv_{\perp,0}
\rangle$ in various coronal holes, fine scale observations of 
magnetic fields with high cadence are required; this is one of the suitable 
targets for Solar-B which is to be launched in 2006.

\begin{acknowledgments}
We thank Profs. Kazunari Shibata, Alex Lazarian, and Bruce Tsurutani for many 
fruitful discussions. 
This work is in part supported by a Grant-in-Aid for the 21st Century COE 
``Center for Diversity and Universality in Physics'' from the Ministry of 
Education, Culture, Sports, Science, and Technology (MEXT) of Japan.  
T.K.S. is supported by the JSPS Research Fellowship for Young
Scientists, grant 4607.
SI is supported by the Grant-in-Aid  (15740118, 16077202)
from the MEXT of Japan.
\end{acknowledgments}


\appendix
\section{Outgoing Boundary Condition}
\label{sc:apd}
We introduce our prescription for the outgoing boundary condition 
of the MHD waves. 
The method is similar to that in \citet{wu01}, while ours is implemented 
for stable treatment of strong MHD shocks. 

Basic MHD equations can be expressed in a matrix form as 
\begin{equation}
\label{eq:apbseq}
\frac{\partial \mbf{U}}{\partial t} + A \frac{\partial \mbf{U}}{\partial r} 
+ \mbf{C} = 0 . 
\end{equation}
\mbf{U} and $A$ are explicitly written as 
$$
\mbf{U} = \left(\begin{array}{l}v_r\\v_{\perp,1}\\v_{\perp,2}\\B_{\perp,1}\\
B_{\perp,2}\\ \rho \\s\end{array}\right), \hspace{0cm}
$$
$$
A = \left( \begin{array}{lllllll} v_r & 0 & 0 & \frac{B_{\perp,1}}{4\pi\rho} & 
\frac{B_{\perp,2}}{4\pi\rho} & c_{\rm s}^2/\rho & p / s\rho \\
0 & v_r &  0 & - \frac{B_r}{4\pi\rho} & 0 & 0 & 0 \\ 
0 & 0 & v_r & 0 &  - \frac{B_r}{4\pi\rho} & 0 & 0 \\
B_{\perp,1} & -B_r & 0 & v_r & 0 & 0 & 0 \\
B_{\perp,2} & 0 & -B_r & 0 & v_r & 0 & 0 \\
\rho & 0 & 0 & 0 & 0 & v_r & 0\\
0 & 0 & 0 & 0 & 0 & 0 & v_r\\
\end{array} \right)  ,
$$
where $s\equiv p/\rho^{\gamma}$ and $c_{\rm s}=\sqrt{\gamma p/\rho}$, and 
subscripts, $\perp,1$ and $\perp,2$, denote first and second transverse 
components. 
$\mbf{C}$ consists of terms due to external force, additional heating and 
cooling, and curvature. In our case, we have taken into account gravity, 
thermal conduction, radiative cooling, and superradial expansion of 
flux tubes:
$$
\mbf{C} = \left(\begin{array}{l}F_1(\frac{B_{\perp}^2}{4\pi\rho} - v_{\perp}^2)
+\frac{G M_{\odot}}{r^2} \\ F_1 (v_r v_{\perp,1} - \frac{B_r B_{\perp,1}}{4\pi 
\rho}) \\ F_1 (v_r v_{\perp,2} - \frac{B_r B_{\perp,2}}{4\pi \rho}) \\ 
F_1 (B_{\perp,1}v_r - B_r v_{\perp,1}) \\ 
F_1 (B_{\perp,2}v_r - B_r v_{\perp,2}) \\ F_2 \rho v_r \\ 
\frac{\rho^{1-\gamma}}{\gamma -1}
(\frac{1}{\rho r^2 f}\frac{\partial}{\partial r}(r^2 f F_{\rm c}) 
+ \frac{q_{\rm R}}{\rho}) \end{array}\right), 
$$ 
where $F_1\equiv \frac{1}{r\sqrt{f}} \frac{\partial}{\partial r} (r\sqrt{f})$ 
and $F_2\equiv \frac{1}{r^2 f} \frac{\partial}{\partial r} (r^2 f)$ 
are arising from the geometrical effect, and $B_{\perp}^2\equiv 
B_{\perp,1}^2 + B_{\perp,2}^2$ and  $v_{\perp}^2\equiv 
v_{\perp,1}^2 + v_{\perp,2}^2$.
Please note that $\mbf{C}$ is negligible except for the thermal conduction 
term ($\frac{1}{\rho r^2 f}\frac{\partial}{\partial r}(r^2 f F_{\rm c})$) 
in our simulations if $r_{\rm out}$ is set at a sufficiently distant location. 

Using eigen values, $\lambda_i$ ($i=1,\cdots ,7$), and eigen vectors, 
$\mbf{I}_i$, defined as 
\begin{equation}
\mbf{I}_i A = \lambda_i \mbf{I}_i, 
\end{equation} 
we can derive characteristic equations for the seven MHD waves from 
eq.(\ref{eq:apbseq}) : 
\begin{equation} 
\mbf{I}_i\frac{\partial \mbf{U}}{\partial t} + \lambda_i \mbf{I}_i
\frac{\partial \mbf{U}}{\partial r} + \mbf{I}_i\cdot \mbf{C} =0
\end{equation}
$\lambda_i$ corresponds to phase speed of each wave, 
$$
\lambda_1 = v_r + v_{\rm f},\hspace{0.1cm}\lambda_2 = v_r + v_{\rm A}, 
\hspace{0.1cm}\lambda_3 = v_r + v_{\rm s}, \hspace{0.1cm}\lambda_4 = v_r, 
$$
\begin{equation}
\lambda_5 = v_r - v_{\rm s},\hspace{0.1cm}\lambda_6 = v_r - v_{\rm A}, 
\hspace{0.1cm}\lambda_7 = v_r - v_{\rm f} ,
\end{equation}
where $v_{\rm f} = \sqrt{\frac{1}{2}[c_{\rm s}^2+\frac{B^2}{4\pi\rho}+ 
\sqrt{(c_{\rm s}^2+\frac{B^2}{4\pi\rho})^2-4 \frac{c_{\rm s}^2 B_x^2}{4\pi
\rho}}]}$, $v_{\rm A}=\frac{B_r}{\sqrt{4\pi\rho}}$, and 
$v_{\rm s} = \sqrt{\frac{1}{2}[c_{\rm s}^2+\frac{B^2}{4\pi\rho} -  
\sqrt{(c_{\rm s}^2+\frac{B^2}{4\pi\rho})^2-4 \frac{c_{\rm s}^2 B_x^2}{4\pi
\rho}}]}$ are phase speeds of fast, \Alfven, and slow modes, respectively. 
Here we define total magnetic field by $B^2=B_r^2+B_{\perp}^2$.  
The eigen vectors are explicitly written as 
$$
\mbf{I}_1 = (\rho v_{\rm f}(v_{\rm f}^2 - v_{\rm A}^2), 
- \frac{B_r B_{\perp,1}}{4\pi} v_{\rm f}, 
- \frac{B_r B_{\perp,2}}{4\pi} v_{\rm f}, 
$$
$$
\frac{B_{\perp,1}^2}{4\pi} v_{\rm f}^2, 
\frac{B_{\perp,2}^2}{4\pi} v_{\rm f}^2, 
c_{\rm s}^2(v_{\rm f}^2 - v_{\rm A}^2), p/s(v_{\rm f}^2 - v_{\rm A}^2))
$$
$$
\mbf{I}_2 = (0,B_{\perp,2},-B_{\perp,1},-\frac{B_{\perp,2}}{\sqrt{4\pi\rho}}, 
\frac{B_{\perp,1}}{\sqrt{4\pi\rho}},0,0)
$$
$$
\mbf{I}_3 = (\rho v_{\rm s}(v_{\rm s}^2 - v_{\rm A}^2), 
- \frac{B_r B_{\perp,1}}{4\pi} v_{\rm s}, 
- \frac{B_r B_{\perp,2}}{4\pi} v_{\rm s}, 
$$
$$
\frac{B_{\perp,1}^2}{4\pi} v_{\rm s}^2, 
\frac{B_{\perp,2}^2}{4\pi} v_{\rm s}^2, 
c_{\rm s}^2(v_{\rm s}^2 - v_{\rm A}^2), p/s(v_{\rm s}^2 - v_{\rm A}^2))
$$
$$
\mbf{I}_4 = (0,0,0,0,0,0,1)
$$
$$
\mbf{I}_5 = (\rho v_{\rm s}(v_{\rm s}^2 - v_{\rm A}^2), 
- \frac{B_r B_{\perp,1}}{4\pi} v_{\rm s}, 
- \frac{B_r B_{\perp,2}}{4\pi} v_{\rm s}, 
$$
$$
-\frac{B_{\perp,1}^2}{4\pi} v_{\rm s}^2, 
-\frac{B_{\perp,2}^2}{4\pi} v_{\rm s}^2, 
-c_{\rm s}^2(v_{\rm s}^2 - v_{\rm A}^2), -p/s(v_{\rm s}^2 - v_{\rm A}^2))
$$
$$
\mbf{I}_6 = (0,B_{\perp,2},-B_{\perp,1},\frac{B_{\perp,2}}{\sqrt{4\pi\rho}}, 
-\frac{B_{\perp,1}}{\sqrt{4\pi\rho}},0,0)
$$
$$
\mbf{I}_7 = (\rho v_{\rm f}(v_{\rm f}^2 - v_{\rm A}^2), 
- \frac{B_r B_{\perp,1}}{4\pi} v_{\rm f}, 
- \frac{B_r B_{\perp,2}}{4\pi} v_{\rm f}, 
$$
$$
-\frac{B_{\perp,1}^2}{4\pi} v_{\rm f}^2, 
-\frac{B_{\perp,2}^2}{4\pi} v_{\rm f}^2, 
-c_{\rm s}^2(v_{\rm f}^2 - v_{\rm A}^2), -p/s(v_{\rm f}^2 - v_{\rm A}^2))
$$

The characteristic equations enable us to implement the outgoing boundary 
condition in a clear and direct manner. 
All the incoming waves should be removed for the proper outgoing boundary. 
This is accomplished by setting physical variables, $\mbf{U}$, to be 
constant in space for the incoming characteristics \citep{tho87}. 
Then, we simply impose following condition at $r=r_{\rm out}$ : 
\begin{equation}
\label{eq:outgo}
\left. \left(\mbf{I}_i \frac{\partial \mbf{U}}{\partial t} + \cal{L}_{\it i}  
+ \mbf{I}_{\it i} \cdot \mbf{C} \right)\right|_{r_{\rm out}} = 0, 
\end{equation}
where 
\begin{equation}
\cal{L}_{\it i} = \left\{ \begin{array}{l@{\quad : \quad}l}
\lambda_i \mbf{I}_i \frac{\partial {\mbox{\boldmath \footnotesize $U$}}}
{\partial r} & \lambda_i > 0 
\:{\rm (Outgoing)} \\
0 & \lambda_i < 0 \: {\rm (Incoming)} \\ 
\end{array} \right.
\end{equation}
 
In our scheme, we firstly perform time evolution to derive 
$\mbf{U}_j^{n}$($1\le j \le j_{\rm out}-1$) at time $n$ except the outermost 
grid, $j_{\rm out}$, from $\mbf{U}_j^{n-1}$($1\le j \le j_{\rm out}$) at 
the previous time step, $n-1$, by the second order MHD Godunov-MOCCT scheme. 
The physical variables 
at the outermost grid, $\mbf{U}_{j_{\rm out}}^n$ are determined at the end 
of each time step by using $\mbf{U}_{j_{\rm out}}^{n-1}$ and 
$\mbf{U}_{j_{\rm out-1}}^{n}$.  
At this time we integrate eq.(\ref{eq:outgo}) implicitly in time for numerical 
stability: 
$$
\hspace{-2cm}
(\mbf{I}_i)_{j_{\rm out}}^n\frac{\mbf{U}_{j_{\rm out}}^n - 
\mbf{U}_{{\it j}_{\rm out}}^{\it n-1}}{\Delta t} 
$$
\begin{equation}
\label{eq:ougopr}
+ (\cal{L}_{\it i})_{{\it j}_{\rm out}}^{\it n} + 
(\mbf{I}_{\it i} \cdot \mbf{C})_{{\it j}_{\rm out}}^{\it n} = {\it 0}. 
\end{equation} 
We adopt upwind discretization for the spatial derivative appeared in 
$\cal{L}_{\it i}$ for the outgoing characteristic, 
$(\frac{\partial {\mbox{\boldmath \footnotesize $U$}}}{\partial r})_{j_{\rm 
out}}^n = \frac{{\mbox{\boldmath \footnotesize $U$}}_{j_{\rm out}}^n 
- {\mbox{\boldmath \footnotesize $U$}}_{j_{\rm out}-1}^n}{\Delta r}$. 
Note that $(\mbf{I}_i)_{j_{\rm out}}^n$, 
$(\cal{L}_{\it i})_{{\it j}_{\rm out}}^{\it n}$, 
and $\mbf{C}_{j_{\rm out}}^n$ are functions of 
$\mbf{U}_{{\it j}_{\rm out}}^n$, so we need iteration to determine 
$\mbf{U}_{{\it j}_{\rm out}}^n$. 
Equation (\ref{eq:ougopr}) determines the physical quantities at the outermost 
mesh to satisfy the outgoing condition for all the seven MHD characteristics.
 
We carry out reflection tests for our implementation by using different types 
of waves with various amplitudes from linear to extremely nonlinear regimes.  
We here consider two types of waves as typical examples : 
(i)\Alfven waves, and (ii) oblique fast waves which propagate 
in 45 degree with respect to the underlying magnetic field.  
We use Cartesian coordinates with homogeneous background for the tests.  
Solitary waves which travel in the right direction are initially set, 
and the outgoing condition is implemented at the right boundary. 
In order to give the waves which purely propagate in one direction even in the 
extremely nonlinear regime, we adopt simple wave solutions. 
Circularly polarized \Alfven waves traveling in 
one direction are the exact solutions of the basic MHD equations.   
For the oblique fast waves, $\rho$, $p$, $v_x$, and $v_y$ can be expressed 
by $B_y$ \citep{wu87}: 
\begin{equation}
\label{eq:smwv1}
\frac{d\rho}{dB_y}=\frac{B_y}{4\pi}\frac{1}{v_{\rm f}^2-c_{\rm s}^2}, 
\end{equation}
\begin{equation}
\frac{d v_x}{dr} = \frac{B_y}{4\pi\rho}\frac{v_{\rm f}}
{v_{\rm f}^2-c_{\rm s}^2}, 
\end{equation}
\begin{equation}
\frac{d v_y}{dB_y} = -\frac{B_x}{4\pi\rho v_{\rm f}}, 
\end{equation}
\begin{equation}
\label{eq:smwv4}
\frac{dp}{dB_y}=\frac{B_y}{4\pi}\frac{c_{\rm s}^2}{v_{\rm f}^2-c_{\rm s}^2}.  
\end{equation}
Note that we can treat fast (and slow) waves in two dimensional 
($x \& y$) space without loss of generality because they are coplanar, while 
the three dimensional space is required for circularly polarized 
\Alfven waves. We initially give sinusoidal variations of $B_y$ for the fast 
waves and the other quantities are 
determined by Equations (\ref{eq:smwv1}) - (\ref{eq:smwv4}). 
If the wave amplitude, $\delta v$, is sufficiently smaller than 
$v_{\rm f}$, all the quantities can be approximated by sinusoidal variations. 

\begin{figure}
\figurenum{15} 
\begin{center}
\noindent\includegraphics[width=20pc]{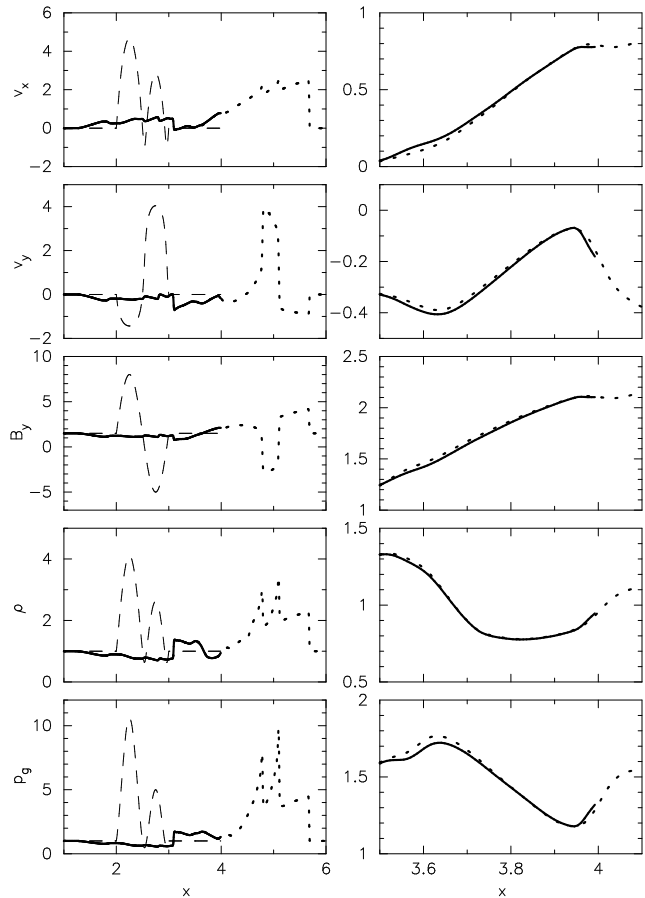}
\end{center}
\caption{Reflection test for the oblique simple fast wave with 
$\delta v/v_{\rm f} = 3.1$.  From top to bottom, $v_x$, $v_y$, $B_y$, $\rho$, 
and $p$ are plotted. In the panels on the right each quantity on the left is 
zoomed-up. Dashed and solid lines denote the initial and final ($t=0.6$) 
states. Dotted lines are results of simulation in the broader region at 
$t=0.6$.  Differences between solid and dotted lines indicate errors due to 
our implementation of the outgoing boundary condition.}
\label{fig:ref2}
\end{figure}

Figure \ref{fig:ref2} shows an example of the reflection tests for 
the oblique fast waves. We non-dimensionalize the $B$ values 
by $1/4\pi \rightarrow 1 $. 
The initial background conditions are, $\rho=1$, $p=1$, $B_x=1.5$, $B_y=1.5$, 
$v_x=0$, and $v_y=0$, which give plasma $\beta$ value, 
$\beta = 2p/B^2$($=8\pi p/B^2$ in cgs-Gauss unit)$=0.44$. 
We give magnetic field amplitude, $\delta B_y 
=6.5$, which corresponds to relatively strong non-linearity, 
$\delta v/v_{\rm f}=3.1$. 
The simulation box is from $x=0$ to $x=4$ and the number of grid points is 
512. 
The initial solitary wave is put between $x=2$ and 3. 
The initial condition ($t=0$) and results at $t=0.6$ are plotted in 
dashed and solid lines. 
To estimate errors at the right boundary, we also carry out simulation 
in larger box between $x=0$ and 8(dotted lines), 
which is free from the boundary errors at $x=4$. 

Because of the strong nonlinearity, the wave rapidly steepens to form shocks 
before reaching $x=4$.  
These shocks travel to the right direction, which are seen in the simulation 
with the larger box size (dotted lines). 
At $t=0.6$, the main part of the wave goes out of the simulation box ($x=4$) 
except the trailing edge.  
The differences between solid and dotted lines indicate errors due to our 
implementation of the outflow boundary condition. 
Figure \ref{fig:ref2} shows the errors are quite small; relative errors 
to the initial amplitude are at most 1\% for each variable. 
Our implementation is good enough for the passage of the moderately 
strong shocks. 

\begin{figure}
\figurenum{16} 
\begin{center}
\noindent\includegraphics[width=20pc]{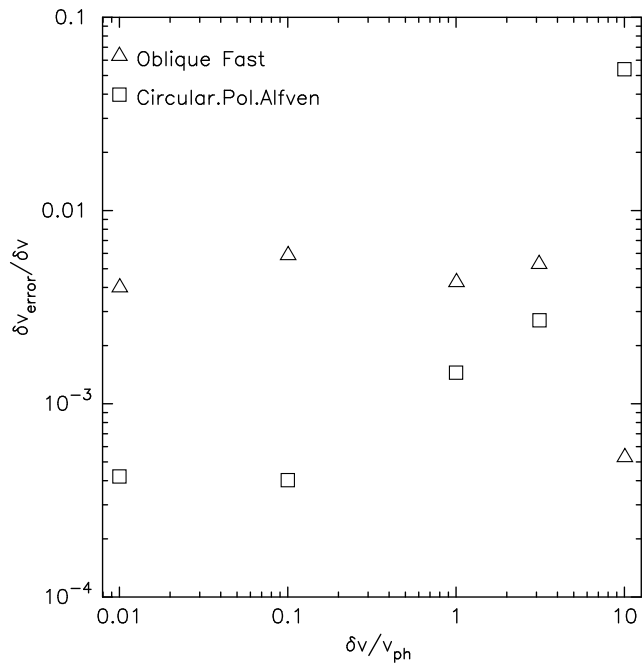}
\end{center}
\caption{Reflection test for the simple fast and the \Alfven waves.
Horizontal axis shows the input wave amplitudes normalized by the phase 
speed. Vertical axis plots errors due to our implementation of the outgoing 
boundary condition normalized by the input amplitude. (see text)}
\label{fig:ref1}
\end{figure}

Figure \ref{fig:ref1} shows the relative errors to the initial amplitude, 
$\delta v_{\rm error}/\delta v$ on 
the normalized wave amplitude, $\delta v/v_{\rm f}$ and 
$\delta v/v_{\rm A}$, for the oblique fast waves and circularly polarized 
\Alfven waves. As for the relative errors, we plot the errors of $v_y$, 
whereas the other quantities ($\rho, p, B_y, v_x$) give similar results. 
If $\delta v/v_{\rm ph}\lesssim 3$ ($v_{\rm ph}=v_{\rm f}$ or $v_{\rm A}$ is
phase speed), the relative errors are smaller than 0.01 and our 
implementation is sufficiently acceptable. 
The \Alfven wave with $\delta v/v_{\rm A}=10$ gives larger 
$\delta v_{\rm error}/\delta v\simeq 0.05$. 
This is because as the wave travels it dose not only consist of \Alfven modes 
but contains non-linear fast and slow wave components.  However, this value 
($\delta v_{\rm error}/\delta v\simeq 0.05$) 
is still acceptable since an error in energy 
($\propto \delta v^2$) is only 0.3\%.

\end{article}

\end{document}